**Title:** Diamond for biosensor applications

# Diamond for biosensor applications

Christoph E. Nebel[a,c)], Bohuslav Rezek[b], Dongchan Shin[a], Hiroshi Uetsuka[a], Nianjun Yang[a]

[a]*Diamond Research Center, AIST, Central 2, Tsukuba 305-8568, Japan*
[b]*Institute of Physics, Academy of Sciences of the Czech Republic, Prague, Czech Republic, Cukrovarnicka 10, CZ-162 53 Praha 6, Czech Republic*

[c)]email: christoph.nebel@aist.go.jp

**Abstract:**

A summary of photo- and electrochemical surface modifications applied on single-crystalline chemical vapor deposition (CVD) diamond films is given. The covalently bonded formation of amine- and phenyl-linker molecule layers are characterized using X-ray photoelectron spectroscopy (XPS), atomic force microscopy (AFM), cyclic voltammetry and field-effect transistor characterization experiments. Amine- and phenyl-layers are very different with respect to formation, growth, thickness and molecule arrangement. We detect a single molecular layer of amine linker-molecules on diamond with a density of about $10^{14}$ cm$^{-2}$ (10 % of carbon bonds). Amine molecules are bonded only on initially H-terminated surface areas to carbon. In case of electrochemical deposition of phenyl-layers, multi-layer formation is detected due to three dimensional (3D) growths. This gives rise to the formation of typically 25 Å thick layers. The electrochemical grafting of boron doped diamond works on H-terminated and oxidized surfaces.

After reacting of such films with heterobifunctional crosslinker-molecules, thiol-modified ss-DNA markers are bonded to the organic system. Application of fluorescence and atomic force microscopy on hybridized DNA films show dense arrangements with densities up to $10^{13}$ cm$^{-2}$. The DNA is tilted by an angle of about 35° with respect to the diamond surface. Shortening the bonding time of thiol-modified ss-DNA to 10 minutes cause a decrease of DNA density to about $10^{12}$ cm$^{-2}$. Application of AFM scratching experiments show threshold removal forces around 75 nN for DNA bonded on phenyl linker-molecules and of about 45 nN for DNA bonded to amine linker-molecules. DNA sensor applications using $Fe(CN)_6^{3-/4-}$ mediator redox-molecules, impedance spectroscopy and DNA-field effect transistor devices performances are introduced and discussed.

# 1 Introduction

Genomics research has elucidated many new biomarkers that have the potential to greatly improve disease diagnostics [1-3]. The availability of multiple biomarkers is important in diagnosis of complex diseases like cancer [4,5]. In addition, different markers will be required to identify different stages of disease pathogenesis to facilitate early detection. The use of multiple markers in healthcare will, however, ultimately depend upon the development of detection techniques that will allow rapid detection of many markers with high selectivity and sensitivity. Currently extensive quests for proper transducer materials, for optimization of detection techniques and sensitivities, for realization of highly integrated sensor arrays, and for bio-interfaces which show high chemical stability which are required in high through-put systems, are therefore on-going. Most of established substrate materials ("transducers") like latex beads, polystyrene, carbon electrodes, gold, and oxidized silicon or glass do not possess all desired properties like flatness, homogeneity, chemical stability, reproducibility and biochemical surface modifications [6-10]. In addition, future technologies require integration of bio-functionalized surfaces with microelectronics or micro-mechanical tools which adds significant complexity to this topic [10-14], as most of microelectronic-compatible materials like silicon, $SiO_x$, and gold show degradation of their bio-interfaces in electrolyte solutions [14].

Diamond can become a promising candidate for bio-electronics as it shows good electronic [15-17] and chemical properties [18-20]. Fig. 1 shows voltammograms for water electrolysis of various electrodes. The supporting electrolyte is 0.5 M $H_2SO_4$. Please note that each current/voltage scan has been shifted vertically for better comparison. Two poly-crystalline films, B:PCD(NRL) with $5x10^{19}$ B/$cm^3$ and B:PCD(USU) with $5x10^{20}$ B/$cm^3$ (from Ref. 21, 22) are compared with a single crystalline boron-doped diamond B:(H)SCD with $3x10^{20}$ B/$cm^3$ and with an undoped diamond (H)SCD [23]. The electrochemical potential-window of diamond is significantly larger and the background current within this regime considerably lower than conventional materials. In addition, by tuning the boron doping level, the onset of hydrogen evolution (rise of current at negative potentials) can be reduced or switched completely off by decreasing the boron doping level from extremely high with $>10^{20}$ $cm^{-3}$ boron ("metallic") to "undoped" (= intrinsic diamond). There are some other parameters affecting the electrochemical potential-window like crystal orientation [24], structural perfection of polycrystalline diamond [25] and surface termination [18]. Their discussion in this context is, however, beyond the scope of this paper.

Surface induced conductivity of hydrogen terminated undoped diamond in electrolyte solutions is another unique property which attracted significant attention in recent years [26]. It is generated by transfer doping of hydrogen-terminated diamond immersed into electrolyte solution. The phrase "transfer doping" indicates that the surface conductivity in diamond arises from missing valence-band electrons as such electrons "transfer" into the electrolyte [27-29]. For such transitions, the chemical potential of an electrolyte must be below the energy level of the valence-band maximum. For most semiconductors this is not the case, as can be seen in Fig. 2. Even for oxidized diamond, chemical potentials are mostly deep in the band-gap of diamond. It changes drastically if the surface of diamond, which consists of about $2x10^{15}$ cm$^{-2}$ carbon bonds, is terminated with hydrogen. Hydrogen-carbon bonds are polar covalent bonds (electronegativity of carbon: 2.5 and of hydrogen: 2.1), therefore a dense surface dipole layer is generated with slightly negative charged carbon (C$^-$) and slightly positive charged hydrogen (H$^+$). From basic electrostatics such a dipole layer causes an electrostatic potential step $\Delta V$ perpendicular to the surface over a distance of the order of the C-H bond length of 1.1 Å. Simple calculations show that the energy variation over this dipole is in the range of 1.6 eV (for a detailed discussion see Ref. 30). This dipole energy shifts all energy levels of diamond for about 1.6 eV up with respect to the chemical potential of an electrolyte (see Fig. 2). Conduction-band states of diamond are shifted above the vacuum level of the electrolyte. This scenario is called "negative electron affinity" (see Fig. 2: clean diamond, where the vacuum level is about 0.3 eV above the conduction band minimum and H-terminated diamond, where the vacuum level is 1.3 eV below the conduction band minimum) [29,31].

As all electronic states are shifted for the same dipole energy, occupied valence-band states emerge above the chemical potential $\mu$ of electrolytes. Electrons from the diamond valence-band (electronically occupied states) can therefore tunnel into empty electronic states of the electrolyte until thermodynamic equilibrium between the Fermi level of diamond and the electrochemical potential of the electrolyte is established. This is schematically shown in Fig. 3a. Fermi-level and chemical potential, μ, align and form a narrow valence-band bending of 20 to 30 Å in width, which is in effect a confined hole accumulation layer [32,33]. Such alignment requires defect free bulk and surface properties as well as a perfect H-termination. During recent years, the growth of diamond has been optimized to such a level in combination with a perfect H-termination of the surface (for reviews see Ref. 34, 35).

Evidence from theory and experiments suggest that the electron affinity of diamond in contact with water is approximately 1 eV more positive than that observed in high vacuum [36-38]. A dominant interaction of diamond energy levels with the $H_2/H^+$ redox states seems to be therefore less likely. But, diamond valence-band states will still scale with interactions to the $O_2/H_2O$ couple, giving rise to the discovered phenomena. As the chemical potential of electrolytes is changing with pH-value, a variation of the surface conductivity can be detected experimentally, following closely the Nernst prediction with 55 mV/pH (see Fig. 3b) [39-41,].

Diamond is known to be biocompatible [42-44] and has therefore a potential for "in-vivo" electronic applications. When Takahashi et al. in 2000 [45,46] firstly introduced a photochemical chlorination/amination/carboxylation process of the initially H-terminated diamond surface, a giant step towards bio-functionalization of diamond was taken, as the obstacle of "chemical inertness" of diamond has finally been removed. This triggered more activities so that two years later, Yang and co-workers in 2002 introduced a new photochemical method to modify nano-crystalline diamond surfaces using alkenes [14], followed by electrochemical reduction of diazonium salts which has been successfully applied to functionalize boron-doped ultrananocrystalline diamond [47] and recently a direct amination of diamond has been introduced [48]. Such functionalized surfaces have been further modified with DNA, enzymes and proteins and characterized using fluorescence microscopy and impedance spectroscopy [14,49,50] voltammetry and gate-potential shifts of ion-sensitive field-effect transistors [51,52].

Maybe the most influential argument for diamond applications in bio-technology has been given by Yang et al. in 2002 [14]. They characterized the bonding stability of DNA to nano-crystalline diamond and other substrates in hybridization/denaturation cycles using fluorescence microscopy investigations. The result is shown in Fig. 4 in comparison to Au, Si and glassy carbon. It demonstrates that DNA bonding to diamond is significantly better than to other substrates as no degradation of fluorescence intensity could be detected. The long-term bonding stability is especially important in multi-array sensor applications which are costly to produce and which therefore need long term stability in high through-put systems.

Applications of diamond sensors will ultimately depend on the commercial availability of diamond films. This has improved significantly during recent years as meanwhile, nano- and polycrystalline CVD diamond films can be grown by plasma enhanced chemical vapor deposition (CVD) heteroepitaxially on silicon and other substrates on large area. Growth parameters are currently optimized to deposit films at low temperature to allow integration into established silicon technology [53-55]. Single crystalline diamond produced by high temperature high pressure growth is commercially available due to an increasing number of companies producing diamond. The size of these layers is relative small, typically 4 mm x 4 mm which is, however, large enough to be use as substrate for homoepitaxial growth of high quality single crystalline CVD diamond ("electronic grade quality").

With respect to electronic applications, a careful selection of "diamond" material is required. Fig. 5 summarizes the structural properties of nano-, poly and single-crystalline diamond. Ultranano-, nano- and polycrystalline diamond layers are dominated by grain-boundaries which are decorated with sp2 and amorphous carbon [56-58]. The volume-fraction of sp2 and grain boundaries depends on growth parameter and varies from layer to layer. Especially ultra-nano-crystalline diamond contains a high volume fraction of up to 5 % [55]. Amorphous carbon and sp2 generate a continuous electronic density-of-states distribution in the gap of diamond. These states will affect sensor sensitivity and dynamic properties. Applications of polycrystalline diamond as photo- or high-energy particle detector show therefore memory and priming effects which arise by metastable filling of grain-boundary states [59,60]. In addition, such diamond films show a significant surface roughness in the rang of 30 to 50 nm for nano-crystalline diamond and micrometer to tens of micrometer for polycrystalline layers. Commercially available polycrystalline diamond is therefore often mechanically polished, to achieve a smooth surface. This generates, however, a thin highly damaged diamond surface which cannot be tolerated in surface related electronic applications as surface defects, about 0.9 to 1.1 eV above the valence band maximum, will pin the Fermi-level and will deteriorate heterojunction properties [61,62].

On the other hand, single crystalline CVD diamond has been optimized over recent years to electronic grade quality with atomically smooth surfaces (see Fig. 5c) [63-65]. These films show even at room temperature strong free-exciton emissions at 5.27 eV and 5.12 eV, which are fingerprints of low defect densities, typically below $10^{15}$ $cm^{-3}$. The bulk resistivity of undoped films at 300 K is larger than $10^{15}$ $\Omega cm$ [56,57]. Atomic force microscopy (AFM) characterization of such films show surface morphologies which indicate atomically flat properties with step-etch growth, where terraces run parallel to the (110) direction. After H-terminated of such layers, heterojunction properties follow very well predicted properties of defect-free diamond.

Diamond can be doped p-type by boron which results in a doping level 360 meV above the valence band maximum [66]. Phosphorus doping has been introduced for n-type doping with the phosphorus doping level 0.6 eV below the conduction band minimum [67]. Both levels are basically too deep for room temperature electronic applications, which is the typical regime for bio-electronics. One way to overcome this problem is the application of metallic doping where in case of boron, typically $10^{20}$ $B/cm^3$ or more atoms are incorporated into diamond [68]. This causes enough wave-function overlap of holes in acceptor states to allow propagation in such states, without thermal activation to the valence-band. Highly-boron doped diamond is therefore well established in electrochemistry. Applications of n-type diamond in electro- or biochemical sensors seem to be not too favourable, as the Fermi-level (0.6 eV below the conduction band) and chemical potential of electrolytes (typically 4.5 eV below the vacuum level (see Fig. 2) are too different, giving rise to energy barrier limited electronic interactions.

This brief introduction of major properties of diamond show, that it is indeed an interesting transducer material for bio-sensor applications. In the following we review our achievements with respect to interface properties of single crystalline CVD diamond to organic linker-molecule layers and DNA films. We describe surface functionalization using amine- and phenyl-layers, which are currently attracting significant attention. There are other photochemical modifications of diamond available (for example see Refs. [45,46] or [48]) which are based on direct or indirect surface amination. It is very likely that sooner or later the modification spectrum will become even broader. In the following, however, we want to focus on i) amine and ii) phenyl-related modifications, as these techniques are established, are used by a growing number of scientists, and are characterized reasonably well.

For our experiments we used homoepitaxial grown, atomically smooth CVD diamond, either undoped or metallically boron doped, which are free of grain boundaries, sp2-carbon or other defects. We apply a) photochemical attachment chemistry of alkene molecules to undoped diamond [69-71] and b) electrochemical reduction of diazonium salts [69,72,73], to form nitrophenyl linker molecules on boron doped CVD diamond. The bonding mechanisms, kinetics, molecule arrangements and densities will be introduced using a variety of experiments like X-ray photoelectron spectroscopy (XPS), scanning electron microscopy (SEM), atomic force microscopy (AFM), cyclic voltammetry and several electronic characterization techniques. By use of a hetero-bifunctional cross-linker, thiol-modified single-stranded probe DNA (ss-DNA) is bonded to diamond. Finally, such surfaces are exposed to fluorescence labeled target ss-DNA to investigate hybridization by use of fluorescence microscopy. We applied AFM in electrolyte solution, to gain information about geometrical properties of DNA, bonding strength, as well as the degree of surface coverage. Finally, we introduce first results with respect to label free electronic sensing of DNA hybridization using $Fe(CN_6)^{3-/4-}$ redox molecules as mediator in amperometric experiments and variation of gate-potential threshold shifts in DNA-FET structures. For a detailed summary of these results see Ref. 69.

## 2.    Materials and Methods

### 2.1    CVD Diamond growth, surface modifications and contact deposition

High quality undoped, single crystalline diamond films of 200 nm thickness have been grown homoepitaxially on 3 mm x 3 mm (100) oriented synthetic Ib substrates, using microwave-plasma chemical vapor deposition (CVD). Growth parameters were: substrate temperature 800 °C, microwave power 750 W, total gas pressure 25 Torr, total gas flow 400 sccm with 0.025 % $CH_4$ in $H_2$. Note, the used substrates have been grown commercially by high pressure high temperature (HPHT) technique and contain typically up to $10^{19}$ cm$^{-3}$ dispersed nitrogen (type Ib diamond). To achieve H-termination after growth, $CH_4$ is switch off and diamond is exposed to a pure hydrogen plasma for 5 minutes with otherwise identical parameters. After switching off the hydrogen plasma, the diamond layer is cooled down to room temperature in $H_2$ atmosphere. A detailed discussion of sample growth properties can be found in Refs. [63-65]. Layers are highly insulating with resistivities larger than $10^{15}$ $\Omega$cm. Surfaces are smooth as characterized by atomic force microscopy (AFM). A typical result is shown in Fig. 6. The root mean square (RMS) surface roughness is below 1 Å.

Boron-doped single-crystalline diamond films have been grown homoepitaxially on synthetic (100) Ib diamond substrates with 4 mm × 4 mm × 0.4 mm size, using microwave plasma-assisted chemical vapor deposition (CVD). Growth parameters are: microwave power 1200 W which generate a substrate temperature around 900 $^o$C, gas pressure 50 Torr, gas flow 400 sccm with 0.6 % $CH_4$ in $H_2$. $B_2H_6$, as boron source, is mixed in $CH_4$, where the boron/carbon atomic ratio (B/C) was 16000 ppm. Typically 1 μm thick films have been grown within 7 hours. H-termination has been achieved in the same way as described above. To measure bulk properties, boron doped diamond is wet-chemically oxidized by boiling in a mixture of $H_2SO_4$ and $HNO_3$ (3:1) at 230 $^o$C for 60 minutes. Fig. 7 shows a typical result of conductivity, σ, which is in the range of 200 $(\Omega cm)^{-1}$ at 300 K, showing a negligible activation energy of 2 meV ("metallic properties"). It is achieved by ultra-high doping of diamond with $3 \times 10^{20}$ $cm^{-3}$ boron acceptors as detected by secondary ion mass spectroscopy (SIMS). The crystal quality is not deteriorated by this high boron incorporation. A series of X-ray diffraction (XRD) and Raman experiments have been applied to investigate details of crystal quality which will be discussed elsewhere.

To obtain patterns of H- and O-termination on diamond surfaces, we apply photolithography, using photoresist as mask to protect H-terminated areas while uncovered surface parts are exposed to a 13.56 MHz RF oxygen plasma. Plasma parameters are: oxygen $(O_2)$ gas pressure 20 Torr, plasma power 300 W, and duration 2.5 minutes. Wetting angle experiments of H-terminated surfaces show angles > 94$^o$ indicating strong hydrophobic properties. After plasma oxidation the wetting angle approaches 0$^o$, as the surface becomes hydrophilic.

For electronic characterization or realization of DNA-field-effect transistors (DNA-FET) we deposited Ohmic contacts on H-terminated diamond by thermal evaporation of 200 nm-thick Au onto photoresist patterned diamond, followed by a lift-off process. In case of highly boron doped diamond Ti (100 Å)/ Pt (100 Å)/ Au (2000 Å) contacts have been realized using e-beam evaporation. Fig. 8 shows a DNA-FET from diamond as measured by scanning electron microscopy (SEM). The sensor area of size 2 mm x 0.7 mm is originally H-terminated diamond which connects drain and source Au-contacts. This area is surrounded by insulating diamond which has been oxidized. The H-terminated surface is chemically modified as described below to covalently bond DNA to it. For experiments in electrolyte solution, drain and source contacts are insulated by silicon rubber. We use Pt as gate electrode (not shown) in buffer solutions.

Electrochemical experiments on boron doped diamond are performed on typical areas of 3 mm$^2$ size. Ohmic contacts to boron doped diamond are evaporated outside of this area and sealed with silicon rubber.

## 2.2    Photochemical surface modification of undoped diamond

Undoped single-crystalline diamond surfaces are modified by photochemical reactions with 10-amino-dec-1-ene molecules protected with trifluoroacetic acid group ("TFAAD") [70,71]. Restricted Hartree-Fock calculation of theoretical geometric properties of TFAAD molecules were performed with the Gaussian 98 package with density functional theory (B3LYP/6-31G(d)) [74]. A typical result is shown in Fig. 9. The length of the molecule is 11.23 Å, and the diameter 5.01 Å. It is interesting to note, that the protecting cap molecule shows a tilted arrangement. In case of up-right arrangement of molecules on diamond, one expects therefore a monolayer thickness of around 11 to 15 Å.

The chemically reactive end of TFAAD is terminated with an olefin (C=C), the other is protected from reactions using a trifluoroacetic cap. Chemical attachment is accomplished by placing 4 micro-liters of TFAAD on the diamond substrate. Then, the TFAAD is homogeneously distributed by spin-coating with 4000 rounds/min in air for 20 s which forms a 5 micrometer thick liquid TFAAD layer. After accomplishment of spin-coating, samples are sealed into a chamber with quartz window in nitrogen atmosphere. Then UV illumination is switched on for a given period of time. The ultraviolet light is generated in a high-pressure mercury lamp with emission at 250 nm of 10 mW/cm$^2$ intensity.

## 2.3    Electrochemical surface functionalization

Electrochemically induced covalent attachment of nitrophenyl molecules has been performed using an Electrochemical Analyzer 900 (CHI instruments), and a three-electrode configuration with a platinum counter electrode and an Ag/Ag$^+$ (0.01 M) reference electrode (BAS, Japan) [72,73]. The active area of the boron doped diamond working electrode is about 0.03 cm$^2$. Electrolyte solution for the reduction of 4-nitrobenzene diazonium tetrafluoroborate is 0.1 M tetrabutylammonium tetrafluoroborate (NBu$_4$BF$_4$) in dehydrated acetonitrile (Wako chemicals, H$_2$O: < 50 ppm). The diazonium salts reduction is performed in a N$_2$-purged glove-box. Nitrophenyl-modified diamond surfaces are then sonicated with acetone and acetonitrile. XPS, AFM and voltammetric experiments have been applied to characterize the surface bonding properties and to reduce the nitrophenyl groups to aminophenyl groups. The nitrophenyl groups grafted on single-crystalline diamond substrate can be considered as covalently bonded free nitrobenzene to diamond as shown in Fig. 10 (Molecular Orbital Calculations).

## 2.4     Hetero-bifunctional cross-linking and DNA attachment

To provide chemically reactive amine groups to the photochemically treated diamond samples, the trifluoroacetamide protecting group was removed by refluxing the TFAAD-modified surface in 2:5 MeOH/H$_2$O with 7 % (w/w) K$_2$CO$_3$.

The electrochemically modified surface of boron doped diamond with nitrophenyl groups (–C$_6$H$_5$NO$_2$) are electrochemically reduced to aminophenyl (–C$_6$H$_5$NH$_2$) in 0.1 M KCl solution of EtOH-H$_2$O to provide reactive aminophenyl groups.

To attach DNA, the amine- or the phenyl-layer is then reacted with 14 nM solution of the heterobifunctional crosslinker sulphosuccinimidyl - 4 - (N - maleimidomethyl) cyclohexane-1-carboxylate in 0.1 M pH 7 triethanolamine (TEA) buffer for 20 minutes at room temperature in a humid chamber. The NHS-ester group in this molecule reacts specifically with the -NH$_2$ groups of the linker molecules to form amide bonds. The maleimide moiety was then reacted with (2 - 4) µl thiol-modified DNA (300 µM thiol DNA in 0.1 M pH 7 TEA buffer) by placing the DNA directly onto the surface in a humid chamber and allow to react for given times between 10 minutes to 12 h at room temperature. As probe ss-DNA we used the sequence S1 (= 5'-HS-C$_6$H$_{12}$-T$_{15}$-GCTTATCGAGCTTTCG-3') and as target ss-DNA the sequence F1 (= 5'FAM-CGAAAGCTCGATAAGC-3'), where FAM indicates the presence of a fluorescence tag of fluorescein phosphoramidite. To investigate mismatched interactions a 4 bases mismatched target ss-DNA (5'-FAM-CGATTGCTCCTTAAGC-3') has been used. For some fluorescence experiments the green label (FAM) has been replace by red fluorescence markers (Cy5). All DNA molecules have been purchased from TOS Tsukuba OligoService (http://www.tos-bio.com, Japan). A schematic summary of chemical modification schemes is shown in Fig. 11.

Denaturation of samples has been performed in 8.3 M urea-solution for 30 minutes at 37 $^o$C, followed by rinsing in deionized water. Samples are then hybridized again for another DNA cycle.

## 2.5    X-ray photoelectron spectroscopy (XPS), atomic force microscopy (AFM) and fluorescence microscopy (FM)

The chemical attachment is characterized using an X-ray photoelectron spectroscopy (XPS) system (Tetra Probe, Thermo VG Scientific) with a monochromatized AlK$\alpha$ source (1486.6 eV) at a base pressure of $10^{-10}$ Torr. Unless otherwise noted, electrons are collected ejected between 25$^o$ and 80$^o$ degree with respect to the surface normal. (atomic sensitivity factors: C, 0.296; F, 1; N, 0.477; O, 0.711). The mean free path of electrons is 36 Å for perpendicular excitation.

Microscopic morphology and structural properties of amine-, nitrophenyl- and DNA-layers have been characterized by atomic force microscopy (AFM) (Molecular Imaging PicoPlus) [75-77]. For DNA characterization, layers were immersed into SSPE buffer (300 mM NaCl, 20 mM NaH$_2$PO$_4$, 2 mM ethylenediaminetetraacetic acid (EDTA), 6.9 mM sodium dodecyl sulphate (SDS), titrated to pH 7.4 by 2M NaOH). The buffer solution enables DNA to assume natural conformation and avoids effects of water meniscus around the AFM tip.

Surface morphologies are investigated in oscillating-mode AFM (O-AFM), where the tip-surface interaction is controlled by adjusting the tip oscillating amplitude to a defined value (AFM set-point ratio measurements) [75-77]. The set-point ratio is defined as $r_{SP} = A_O/A_{SP}$, where $A_O$ is the amplitude of free cantilever oscillations and $A_{SP}$ the amplitude of the tip, approaching the surface. Measurements are made typically with $A_O$ of 6 and 10 nm. In addition, we used also cantilever phase shift detection (phase lag of cantilever oscillation with respect to oscillation of the excitation piezo-element) to enhance the material contrast between diamond and DNA.

Molecule bonding properties (mechanical properties) of linker and DNA layers have been characterized by contact mode AFM where we applied different loading forces to the AFM tip (C-AFM) in the range 6 to 200 nN. The scan rate was (10 - 20) μm/s. For forces above a critical threshold, linker and DNA molecules are removed. The difference in height is then measured in O-AFM. Doped silicon AFM cantilevers are used in these experiments with a spring constant of 3.5 N/m. The cantilever resonance frequency is 75 kHz in air and 30 kHz in buffer solution.

Fluorescence microscopy has been applied using a Leica Fluorescence Imaging System DM6000B/FW4000TZ where the fluorescence intensity is evaluated by grey-scale analysis (Leica QWin software). Please note, that we have characterized all diamond layers before surface modifications to detect fluorescence emission arising form the bulk of diamond, like for example from nitrogen/carbon-vacancy complexes. Those samples have been excluded from our experiments. The shown fluorescence is therefore truly from fluorescence-labeled DNA.

## 2.6 DNA-Field effect transistors

To realize in-plane gate DNA field-effect transistors (DNA-FET), undoped CVD diamonds with atomically smooth surfaces have been grown by microwave plasma assisted chemical vapor deposition (see above) on Ib substrates. The layer thickness is typically 200 nm. After hydrogen-termination of the diamond surface, an H-terminated sensor area of 2 mm x 0.7 mm size has been processed by photolithography and plasma oxidation. Two Au contacts (0.7 mm x 0.5 mm) evaporated to each end of the H-terminated surface serve as drain and source contacts (see Fig. 8).

Alkene cross-linker molecules are then attached by photochemical means for (12 - 20) h, followed by the attachment of probe ss-DNA. The density of ss-DNA has been varied between $10^{12}$ and $10^{13}$ cm$^{-2}$ for these experiments. Samples are then transferred into a polyetheretherketon (PEEK) sample holders as shown in Fig. 12a. Please note, that during the transfer the ss-DNA layer is covered by sodium chloride buffer solution (1M with 0.1 M phosphate pH 7.2). The drain/source contacts are sealed against contact with buffer solution by a silicon rubber (see Fig. 12b) which is also used to press platinum wires to the drain and source Au contacts (Fig. 12a). A top view of the set-up is shown in Fig. 12c. The PEEK top part has been designed in a way that 1mm of the sensor area is exposed to buffer solution. To apply well defined gate potentials we use a thin Pt wire of 0.2 mm diameter which is immersed into the sodium-chloride buffer. Drain source currents are measured as function of gate potential that has been varied between 0 and -0.6 V. The gate threshold potential of DNA-FETs has been characterized applying several cycles where firstly properties of the FET with probe ss-DNA have been determined, followed by determination of FET properties after hybridization with complementary target ss-DNA. Then the sample has been denatured and characterized again.

## 3. Results

### 3.1 Photochemical surface modifications of undoped single crystalline diamond

The H-terminated samples are photochemically reacted with long-chain $\omega$-unsaturated amine, 10-aminodec-1-ene, that has been protected with the trifluoroacetamide functional group [69-71]. We refer to this protected amine as TFAAD. A variety of attachment experiments on H-terminated and oxidize diamond surfaces show that this process only works on H-terminated diamond.

To characterize the molecule arrangement and layer formation of TFAAD on hydrogen terminated CVD diamond, we applied firstly AFM scratching experiments with the aim to 1) identify the threshold force for mechanical removal of TFAAD molecules, and 2) to use the step-edge of scratched areas, to measure the thickness of the TFAAD layers. Please note that diamond is ultra-hard (100 GPa hardness), the AFM cantilever will therefore not penetrate diamond even at highest applied loading forces. Such contact mode AFM experiments carve-out rectangular trenches in the amine layer. AFM tapping-mode line scans are then used to measure the height profile across scratched areas.

Forces below 100 nN remove only a fraction of TFAAD molecules while forces equal or larger than 100 nN give rise to complete removal (for details see Ref. 77). A typical result of AFM characterization applied to an amine-layer which has been photochemically attached for 1.5 h is shown in Fig. 13. The TFAAD molecules were removed by 200 nN applied loading force from an area of $2 \times 2 \ \mu m^2$ to truly remove the amine layer. The upper image shows topography and the lower image a typical line profile. Within the cleaned area, the surface roughness of diamond is in the range 1.2 to 2.3 Å. On the amine film we detect a roughness between 3.8 to 9.3 Å. The TFAAD-layer thickness is determined to be in the range 5 to 13 Å which indicates dispersed layer properties as for a closely packed, dense layer the height is expected to be around 11 Å. The variation of film properties as a function of UV-photochemical attachment times is shown in Fig. 14. For times shorter than 4 hours, the average layer thickness increased from zero (0 h) to 9.25 (1.5 h) and 13.3 Å (4h), which is about the length of up-right standing TFAAD molecules on diamond. Illumination time between 4 and 10 hours do not result in a remarkable enlargement of the layer thickness, which saturates around 14 Å, but the average roughness is decreasing. These films are closed with no structural defects like pin-holes in it. Illumination longer than 10 hours then gives rise to further enlargement, the film becomes thicker, for example 31.2 Å after 20 hours, which is about three times the length of TFAAD molecules. Notice, the bars in Fig. 14 indicate an average roughness of films. After short term attachment (< 10 h) the roughness is large (about ±4.0 Å) while layers attached for 10 to 12 h are relative smooth. Longer attachment times again give rise to enlarged roughness (±8.0 Å after 20 h). We attribute this to: a) 2D-formation of a dispersed sub-molecular layer for times < 10 h, b) 2D-formation of a dense mono-molecular layer for times between 10 and 12 hours and c) cross polymerization and 3 D growth for times longer then 12 h.

To investigate the chemical bonding of TFAAD molecules to diamond X-ray photoelectron spectroscopic measurements (XPS) have been applied. Fig. 15a shows XPS survey spectra of a clean hydrogen-terminated single crystalline diamond surface before and after exposed to TFAAD and 10 mW/cm$^2$ UV illumination intensity (254 nm) for 2 hours. Before XPS measurements, samples were rinsed in chloroform and methanol (each 5 minutes in ultrasonic). The overall spectrum shows a strong fluorine peak with a binding energy of 689 eV, an O(1s) peak at 531 eV, a N(1s) peak at 400 eV and a large C(1s) bulk peak at 284.5 eV. Please note that on clean H-terminated diamond no oxygen peak can be detected. The C(1s) spectrum reveals two additional small peaks at 292.9 eV and 288.5 eV (see Fig. 15b) which are attributed to carbon atoms in the CF$_3$ cap group and in the C=O group, respectively. From these experiments, we conclude that UV light of about 250 nm (5 eV) initiates the attachment of TFAAD to H-terminated single crystalline diamond. The ratio of the F(1s) XPS signal (peak area) to that of the total C(1s) signal ($R_{FC}$) as a function illumination time is shown in Fig. 15c. The time dependence of $R_{FC}$ follows approximately an exponential law: $R_{FC} = A\{1-\exp(-t/\tau)\}$, with a characteristic time constant $\tau$ of 1.7 h. Saturation of the area ratio F(1s)/C(1s) is achieved after about 7 h.

Angle resolved XPS experiments shown in Fig. 15d are used to calculate the density of bonded TFFAD molecules to diamond in absolute units. As parameters we used the following data: density of carbon atoms = $1.77 \times 10^{23}$ atoms/cm$^3$, atomic sensitivity factors for C (0.296), and for F (1), and a mean free path for perpendicular illumination = 36.7 Å. Taking into account the area ration F(1s)/C(1s) results in about $2 \times 10^{15}$ cm$^{-2}$ TFAAD molecules bonded after 7 h. This corresponds to the formation of a mono-layer TFAAD as the surface density of carbon bonds on diamond is $1.5 \times 10^{15}$ cm$^{-2}$. However, the TFFAD layer itself consists out of 12 carbon atoms which contribute to the signal, so that the real coverage is smaller than this number (see discussion below).

Fig. 16 shows a combination of AFM thickness data (Fig. 14) and XPS results Fig. 15c. The combined data support our three-step formation model of TFAAD film growth on diamond which is governed by: a) Formation of a sub-mono-molecular layer (t < 8h). b) Formation of single-molecular layer (8h < t <12h). c) Slow, but continuous multi-layer formation by cross-polymerization (t > 12 h).

A perfectly smooth surface of (100) oriented diamond contains about $1.5 \times 10^{15}$ cm$^{-2}$ carbon bonds which are terminated by hydrogen as shown schematically in Fig. 17 [78]. The diameter of the 10-amino-dec-1-ene molecules (not taking into account the trifluoroacetic acid top) is 5.01 Å so that in case of a closely packed TFAAD film the up-right standing molecule will require an area of about $2 \times 10^{-15}$ cm$^{-2}$ (if we assume rotational symmetry). This gives a closed layer density of $5 \times 10^{14}$ cm$^{-2}$. Each TFAAD molecule covers the area of 6 hydrogen atoms. Only one of those hydrogen bonds needs to be broken to bond an amine molecule. Therefore most of the diamond surface may be still H-terminated after generation of a densely packed TFAAD layer.

As hydrogen cannot be detected by XPS we applied an additional experiment to investigate if the surface is still terminated with hydrogen. It is known that H-terminated undoped CVD diamond shows a surface conductivity in buffer solution which arises from transfer doping [29,34,35]. We have therefore measured the drain-source current variation before and after TFAAD attachment using a typical ISFET geometrical arrangement in SSPE buffer solution. Fig. 18 shows the conductivity of a perfect H-terminated single crystal diamond before (open circles) and after (open squares) photochemical attachment of TFAAD for 20 h in SSPE buffer solution. A perfectly (100%) H-terminated diamond gives rise to a drain–source current of approximately 7.5 µA at $U_G$ = -0.6 V. After photoattachment of TFAAD molecules for 20 h, the drain–source current decreased to 3 µA, which is 40% of the initial drain–source conductivity. In the case of 100% H-removal by photo-attachment, the drain–source conductivity would completely disappear. As the wetting properties of TFAAD covered diamond surfaces are also changing, which may affect transfer-doping properties of the surface, we leave a quantitative interpretation of this result to further research activities. The highest theoretical packing density of TFAAD of $5 \times 10^{14}$ $cm^{-2}$ would require the removal of only 17 % of all hydrogen atoms terminating the surface. Following recently published data in the literature the packing density is most likely less than this number and in the range of $2 \times 10^{14}$ $cm^{-2}$ [71,77,79]. Our experimental data as well as these theoretical considerations indicate that the diamond TFAAD interface is still reasonably well terminated with hydrogen. This is promising with respect to sensor applications as high sensitivity will require a defect free interface.

Detailed in-situ characterization of the attachment process indicates that electron emission by sub-bandgap light triggers the covalent bonding of TFAAD molecules to diamond [70,71]. In this process, valence-band electrons are optically excited into empty hydrogen-induced states slightly above the vacuum level as shown schematically in Fig. 19 [30]. From there they can reach unoccupied π* states of TFAAD molecules, generating a nucleophilic situation in the C=C bonding structure.

These radical anions may abstract hydrogen from the surface as shown schematically in Fig. 20, creating a carbon dangling-bond at the surface which itself is very reactive towards olefins. To obtain covalently bonded ordered mono-molecular layers on diamond requires however some additional features. In case of random self-assembly of molecules on diamond, a chaotically organized layer would be generated as the basic requirement of surface mobility to allow intermolecular forces to play their ordering role is missing. Such a process would resembles a "dart game" since the grafted moieties would be irreversibly immobilized on the surface due to the formation of strong covalent C-C bonds. In our case ordered formation of mono-layers are detected which is a strong argument in favor of the model of Cicero et al. [80]. In their investigation of olefin addition to H-terminated silicon they deduce a model which requires the formation of alkene anions that abstract H atoms from the surface, thereby creating carbon surface dangling bonds that covalently bond to other alkene molecules in the liquid. Surface dangling bonds are reactive towards alkenes as demonstrated by Cicero et al. [80]. They showed that the olefin addition on H-terminated Si (111) surfaces follows a chain reaction, initiated at isolated Si dangling bonds. This surface dangling bonds further react with olefins to for carbon centered radical bonded to diamond. These radicals abstract hydrogen from neighboring H-C bonds and thereby regenerating surface dangling bonds which then propagate the reaction as depicted schematically in Fig. 20. Such an ordered and self-organized layer formation seems to be a very reasonable model for our findings. The surface of grafted diamonds after 10 h of attachment appear closed with no pinholes. Some modulation is however obvious which may reflect the fast that the ordered growth will break-down and compete with other spots where the same process has been triggered. One can expect therefore some disorder due to competing domain growth.

The number of electrons photo-excited from diamond into the olefin film is huge as only one out of about 1500 triggers chemical bonding to diamond [71]. These electrons create radical anions which give rise to cross polymerization and particle formation [79]. The cross polymerization on the monomolecular amine layer bonded to diamond is however slow and may arise from the fact that the amine layer on diamond prevents the generation of radical anions as TFAAD molecules cannot approach the surface for a electron transition. The radical anions are generated predominantly on diamond surface areas which are not grafted. These radicals then need to diffuse, driven by statistical properties, to find reaction partners and to cross-polymerize. To bond to immobilized TFAAD molecules seems less likely than to react with other molecules in the near vicinity in liquid phase.

## 3.2 DNA bonding and geometrical properties

To attach DNA, we applied the recipe introduced by Yang et al. 2002 [14], and Hamers et al. 2004 [81], where the protected amine is firstly deprotected, leaving behind a primary amine. The primary amine is then reacted with the heterobifunctional crosslinker sulphosuccinimidy l-4- (N- maleimidomethyl) cyclohexane-1-carboxylate and finally reacted with thiol-modified DNA to produce the DNA-modified diamond surface. In our experiments we use a 4 μl droplet which is placed on the diamond layer and which covers a circular area of about 2 mm in diameter, thereby covering oxidized and H-terminated surface parts. To assess whether DNA-modified diamond surfaces have been generated such surfaces have been exposed to complementary oligonucleotides that were labeled with fluorescence tags FAM. Fig. 21 shows the result of intense green fluorescence (= 100 %) from originally H-terminated regions and less intense fluorescence ($\cong$ 70 %) from oxidized surface areas. The weak fluorescence contrast has two reasons. The first is that non-covalently bonded DNA is attached to oxidized diamond. This will be discussed in the following, using atomic force microscopy experiments (AFM). The other is that transparent diamond gives rise to light trapping so that the transparent diamond appears green in case of fluorescence emission.

### 3.3. Atomic force characterization (AFM) of DNA in single crystalline diamond

DNA-functionalized and hybridized surfaces are characterized by AFM measurements in 2x SSPE/0.2% SDS (sodium dodecyl sulphate) buffer solution [75,82]. By performing contact mode AFM scratching experiments, DNA can be removed and at the interface between clean and DNA covered diamond, the height of DNA has be measured. In addition, the force required to penetrate and remove DNA has be determined, giving insight into the mechanical stability of the bonding. Scratching experiments were performed with different tip loading forces between 10 and 200 nN. A typical result of such experiments on a diamond surface modified with double-strand (ds) DNA is shown in Fig. 22a. For each force, an area of 2 μm x 10 μm has been scratched (scan rate 20 μm/s). Forces around 45 nN (±12 nN) generate a surface which appears to be clean of DNA, as also detected by fluorescence microscopy shown in Fig. 22b.

C-AFM experiments on the boundary between initially oxidized and H-terminated diamond are shown in Fig. 23. With C-AFM, we detect non-covalently bonded DNA on oxidized diamond. These molecules can be removed with forces around 5 nN. This is 5 times lower than DNA bonded to the H-terminated surface. The layer is also significantly thinner as on H-terminated diamond.

By measuring across the boundary between the DNA-functionalized and the cleaned surface, using O-AFM, the DNA layer thickness can be obtained as shown in Fig. 24. O-AFM is preferable to C-AFM on soft layers as the tip-surface interaction can be minimized by monitoring the phase shift of the cantilever oscillations [83]. The phase shift was measured as a function of the set-point ratio, $r_{sp} = A_o/A_{sp}$, where $A_{sp}$ is the set-point amplitude and $A_o$ is the amplitude of free cantilever oscillation, on DNA-functionalized and cleaned diamond surface regions (see squares in Fig. 24a). Fig. 25 summarizes results of phase contrast and set-point ratio measurements. The phase contrast between diamond and DNA is positive and approaches zero for set-point ratios approaching 1, i.e. for increasing tip-surface distance. For a phase contrast near zero, which corresponds to minimized tip-surface interaction, the DNA layer thickness reaches about 75 to 78 Å. Simple O-AFM measurements results in about 70 Å (see Fig. 24) which is slightly smaller than the real DNA thickness. The height of 75 to 78 Å is, however, still significantly lower than the expected height of about 130 Å for up-right standing DNA (105 Å) on linker (12 Å) and cross linker molecules (6 Å) and fluorescence marker FAM6 (5 Å). We attribute this to a tilted arrangement of DNA molecules as shown in Fig. 26 (schematic view of molecules). Using triangular geometry, the tilt angle is around 33º to 36º, which is similar to results of DNA bonded to gold [82,83]. There, the detected film thickness of DNA layers has been attributed to orientational properties of DNA duplexes in the monolayer. Due to the large charge density of the DNA backbone (2-/base pair without condensed counter-ions), the orientation of the individual helices is very sensible towards neighboring molecules and to surface charges. As a consequence, small changes in applied electrochemical potential can cause drastic changes in helical orientation (for a summary, see ref. 82). In our case, the tilted average helical packing orientation arrangement of 35º certainly reflects the minimization of Coulomb repulsion forces between individual duplexes. Effects from diamond surface charges (C-H dipole) or from externally applied electric fields on DNA arrangements are currently investigated by variations of buffer solution ionicities and by application of externally applied electric fields to the diamond transducer.

A topographic surface profile of DNA double helix molecules, bonded on diamond is shown in Fig. 27. It reveals broad undulations due to collective interaction of several

DNA oligomers with the tip. The height is modulated with a periodicity of about 30 to 50 nm. The DNA surface roughness is around ± 5 Å. No pinholes can be detected in the layer. Obviously a closed DNA film has been synthesized on diamond using photochemical attachment.

For sensor applications, dilute DNA films in the range $10^{12}$ to $10^{13}$ cm$^{-2}$ are required [84]. To decrease the bonding density we reduce the time of marker DNA attachment. A saturated and very dense film is achieved after 12 h exposure. Following the arguments of Takahashi et al. 2003 [45,46] this will result in a DNA density of about $10^{13}$ cm$^{-2}$. By decreasing the time of marker ss-DNA attachment, the density can be decreased to $10^{12}$ cm$^{-2}$ as shown in Fig. 28. Here, we have evaluated the change in fluorescence intensity after hybridization, where the attachment of ss-DNA maker molecules has been varied between 10 minutes to 12 h. The attachment kinetics is well described empirically by an exponential function with a time constant, τ, of about 2 h.

## 3.4 Electrochemical surface modifications of boron doped single crystalline diamond

In the following we summarize the electrochemical modification of highly conductive p-type single crystalline CVD diamond [72,73,76]. The general chemical scheme is shown in Fig. 11 (I) where hydrogen-terminated or oxidized diamond is modified with phenyl molecules (step a) to b)). Figure 29 shows cyclic voltammograms of 4-nitrobenzene diazonium salts (1 mM) reactions on highly B-doped single-crystalline CVD diamond film in 0.1 M NBu$_4$BF$_4$ acetonitrile solution (san rate: 0.2 V/s). An irreversible cathodic peak of the first sweep at -0.17 V (vs. Ag/AgCl) indicates nitrophenyl group attachment by diazonium salt reduction [72,73,76]. The reduction peak on single crystalline diamond films decreases rapidly with increasing number of scans within +0.5 to -1.0 V (vs. Ag/AgCl), due to increasing surface passivation with nitrophenyl molecules.

The electrochemical attachment works on H-terminated diamond but also on oxidized diamond surfaces as can be seen in Fig. 30. The voltammetric peak is shifted from -0.17 V for H-termination to -0.41 V (vs. Ag/AgCl) for oxidized diamond. This potential shift of $\Delta U = 240$ mV arises by a change of heterojunction properties at the solid/liquid interface. One significant change arises by the variation of the electron affinity as H-terminated diamond has a negative electron affinity of around -1.1 eV, while oxidized diamond shows positive affinities [31,61]. In addition, it is known that oxidized diamond surfaces show surface Fermi-level pinning due to surface defects [85]. The termination of carbon bonds with oxygen instead of hydrogen adds additional complexity as the bonding energies are changing from C-H of 413 kJ/mol to 360 kJ/mol for C-O and to 805 kJ/mol for C=O, respectively.

After electrochemical derivatization, the diamond substrates are sonificated in acetonitrile, acetone, and isopropanol, in order to investigate properties of attached nitrophenyl layers in 0.1 M NBu$_4$BF$_4$ solution. Nitrophenyl groups grafted on single-crystalline diamond films show two reversible electron transfer steps, which are reproducibly detected in all potential sweeps (see Fig. 31). The generalized reversible redox reactions of this system are summarized in Fig. 32. To estimate the surface coverage ($\Gamma$) of nitrophenyl groups on diamond, we use the transferred charge of the electron transfer reaction at -1.17 V (see shaded area in Fig. 31). This results in $3.8 \times 10^{-7}$ C/cm$^2$ or $8 \times 10^{13}$ molecules/cm$^2$, indicating the formation of a sub-monolayer on diamond (5 % coverage) [72,76].

Detailed investigations of phenyl-layer formation on other electrodes show that this simple interpretation is misleading [86,87]. In most cases multi-layers are deposited. We have therefore applied additional experiments to characterize the growth and thickness of the phenyl-layer on diamond [76]. Generally, all performed contact- and oscillatory-mode AFM experiments on deposited nitrophenyl layers reveal layer thicknesses in the range 28 to 68 Å. Taking into account the length of nitrophenyl molecules of about 8 Å (Fig. 10), it indicates clearly multi-layer formation by cyclic voltammetry deposition.

A better controlled growth of phenyl films on diamond can be achieved by electrochemical means, applying fixed potentials for a given period of time instead of potential cycles [87]. In our case we applied -0.2 V (vs. Ag/AgCl) and measured the transient current during the deposition. The result is shown in Fig. 33. In case of unlimited electron transfer and diffusion limited attachment the dynamics follow the Cottrell law (i(t) $\approx$ t$^{-0.5}$) as introduced by Allongue et al. in 2003 [87]. At the very beginning of the transient current, such a characteristic can be detected. However, for longer times, the current decays faster then predicted by this law. We assume that the growing phenyl-layer limits an effective electron transfer, slowing down the bonding process. The density of electrons involved in this reaction saturates around $4x10^{15}$ cm$^{-2}$.

To verify the layer formation, contact mode AFM has been applied. With increasing force to the tip, the phenyl-layer can be removed from diamond. A typical result is shown in Fig. 34, where the nitrophenyl layer has been attached by one cyclic voltammetry scan (from +0.5 V to -1.0 V vs. Ag/AgCl at a scan rate of 200 mV/s). Forces below 100 nN do not damage to the phenyl film. Above 100 nN, a layer of 26 Å thickness is removed and forces above 120 nN give rise to the removal of a thin layer, 8 Å in height. We assume that firstly a random oriented phenyl layer is removed, while forces above 120 nN are required to remove phenyl linker molecules bonded to diamond.

AFM characterization on phenyl layers, which have been grown at a constant potential of -0.2 V for different times, indicate 3D-growth as shown in Fig. 35 (applied tip force: 200 nN, scan rate: 4 µm/s). After short time attachment (5 s) the thickness of the layer is already between 8 and 23 Å. The thickness variation decreases with increasing attachment time, saturating around 25 Å. Taking into account the saturated electron density of $4x10^{15}$ cm$^{-2}$ and the final thickness of the phenyl-layer of 25 Å, the phenyl molecule density in the layer is about $2x10^{21}$ cm$^{-3}$.

The orientation of phenyl molecules has been characterized by angle resolved XPS experiments. The integrated peak intensities of O(1s) to C(1s) shows a strong angle dependence for attachments at -0.2 V for times up to 40 s (see Fig. 36). We attribute this to an oriented growth of nitrophenyl, with $NO_2$ molecules preferentially located on the growing top of the layer. This is different in case of much thicker layers (30 to 65 Å), attached by 5 cycles in the range +0.5 V to -1.0 V vs. Ag/AgCl at a scan rate of 200 mV/s. Here the XPS angle-variation is weak; molecules are arranged in a more disordered structure.

From these experiments we conclude that the formation of phenyl layers on diamond is governed by 3D growth, with preferential alignment of $NO_2$ cap molecules on the top of growing films, if films are not growing too thick. Growth saturates at a layer thickness of about 25 Å, using constant potential attachment (-0.2 V), while significantly thicker layers in the range 35 to 65 Å are detected after cyclic attachment (+0.5 V to -1 V). Properties of such thick layers are governed by a more random molecule orientation. This is schematically summarized in Fig. 37.

Subsequently, nitrophenyl groups are electrochemically reduced to aminophenyl (–$C_6H_5NH_2$) in 0.1 M KCl solution of EtOH-$H_2O$ solvent (see Fig. 38). The first voltammetry sweep gives rise to an irreversible reduction peak at -0.94 (vs. SCE), which is not detected in the 2nd and higher cyclic voltammetry cycles. Such modified surfaces are then used for chemical bonding to hetero bifunctional cross-linker molecules, SSMCC and thiol-modified probe DNA oligonucleotides.

Figure 39 shows a fluorescence image of S1 ss-DNA marker molecules bonded to diamond after hybridization with its complementary ss-DNA target molecules F1 labeled with Cy5. The image shows DNA bonding to initially H-terminated diamond and to oxidized diamond. The laid "T" shape pattern in Fig. 39 arises from surface oxidation. The fluorescence from this area is about 10 % darker than from hydrogen terminated diamond. As the light intensity is proportional to the density of fluorescence centers, the density of DNA bonded to oxidized diamond is about 10 % smaller than on H-terminated diamond. No fluorescence can be detected using a 4-base mismatched ss-DNA target molecule for hybridization

## 3.5 Atomic force characterization of DNA bonded electrochemically to boron doped single crystalline diamond

Geometrical properties as well as density and bonding strength of DNA bonded by this electrochemical technique to boron-doped diamond has been characterized by AFM experiments as described in paragraph 2.5 and 3.3.

The height of DNA layers is detected to be around 90 Å. This is slightly higher then in case of photo-attachment and arises from the thicker linker molecule layer which is about 25 Å in case of phenyl and 12 Å for amine linkers. Again a tilted arrangement is deduced, comparable to results from photoattachment ($\approx 35°$). The layer is dense with no pin-holes. The removal forces are between 60 and 122 nN, the statistical average around 76 nN. It is interesting to note that on initially oxidized diamond the forces are lower, namely in the range of 34 nN. A comparison of forces is shown in Fig. 40. These results indicate strong bonding of DNA to diamond for both, photo- and electrochemical surface modifications. Removal forces are about two times higher then detected on Au and Mica as summarized in Fig. 41 [88-91]. This is promising with respect to diamond bio-sensor applications where exceptional chemical stability is required.

### 3.6 Electronic detection of DNA hybridization

Most of DNA detection techniques are based on DNA hybridization events. In DNA hybridization, the target ss-DNA is identified by a probe ss-DNA which gives rise to hybridization. This reaction is known to be highly efficient and extremely specific. Commonly used DNA detection techniques (radiochemical, enzymatic, fluorescent) are based on the detection of various labels or reagents and have been proven to be time-consuming, expensive and complex to implement. For fast, simple and inexpensive detection, direct methods are required. In the following we want to introduce recent achieved results with respect to field effect and voltammetric sensing, using single crystalline CVD diamond as transducer.

### 3.6a DNA-FET

Diamond ion-sensitive field effect transistors show sensitively variation of pH with about 55 mV/pH [39-41]. This is close to the Nernst limit of 60 mV/pH. The effect arises from transfer doping so that no gate-insulator layer is required. The separation of surface channel and electrolyte is therefore very small. The application of such a FET system for DNA hybridization detection is new and we show in the following some typical results, achieved on structures with $10^{12}$ cm$^{-2}$ to $10^{13}$ cm$^{-2}$ ss-DNA markers bonded to diamond. A schematic figure of the device with hybridized ds-DNA is shown in Fig. 42. We used a 1M NaCl solution (containing 0.1 M phosphate with pH 7.2) with a Debye length of 3 Å as calculated by [84]

$$\lambda_D = \left( \frac{\varepsilon_{el}\varepsilon_o kT}{2z^2 q^2 I} \right)^{1/2}$$

where k is the Boltzmann constant, T the absolute temperature, $\varepsilon_o$ the permittivity of vacuum $\varepsilon_{el}$ the dielectric constant of the electrolyte, z is the valency of ions in the electrolyte, q is the elementary charge and I represents the ionic strength, for a 1-1 salt, it can be replaced by the electrolyte concentration $n_o$. As the linker molecule is 10 to 15 Å (amine) and the cross-linker molecule 5 Å long, the DNA is not in touch with the Helmholz layer in our experiments.

Fig. 43a shows a comparison of drain source currents ($I_{DS}$) measured at a fixed drain-source potential of -0.5 V as function of gate potential for ss-DNA marker molecules attached on the gate, for hybridized marker and target DNA on the gate and after removal of DNA from diamond. The initial ss-DNA density bonded to diamond is $4x10^{12}$ cm$^{-2}$. The drain source current increases by hybridization as detected also for a sensor where the initial ss-DNA density is slightly smaller ($10^{12}$ cm$^{-2}$) or larger ($10^{13}$ cm$^{-2}$). The gate potential variations from ss-DNA to complementary ds-DNA bonding vary as shown in Fig. 43b between 30 to 100 mV. There is a clear trend that with decreasing DNA-density, the potential shift becomes larger (as predicted by Poghossian et al. 2005 [84]). Taking into account the ion-sensitivity of diamond ISFETs of 55 mV/pH, this reflects a decrease of pH of the buffer solution of about 1 to 1.4 by hybridization. Poghossian et al. 2005 [84] have calculated that the average ion concentration within the intermolecular spaces after hybridization can be more than three to four times higher for cations than before hybridization. In case of a Nernstian slope of the sensor, they predict a gate potential shift of 28 to 35 mV. Our results indicate a stronger change. The increase in drain-source current with hybridization can be well described by the transfer doping model as the increase in cation density will cause a decrease of pH. Therefore the chemical potential will increase giving rise to enhanced surface conductivity.

A summary of published sensitivities of DNA FET's from silicon is shown in Fig. 44 (data from [84,92,93]. The initial sensitivities reported before 2004 of more than 100 meV threshold-potential shifts by hybridization have not been reproduced. On the contrary, sensitivities are decreasing towards experimental reproducible as well as theoretical predictable values in the range 30 to 80 meV.

## 3.6b Cyclic voltammetry and impedance spectroscopy

For amperometric detection of DNA hybridization we use $Fe(CN_6)^{3-/4-}$ as mediator redox molecule [94]. The detection principle is shown in Fig. 45. In case of ss-DNA bonded to diamond (Fig. 45a), the relative small negatively charged redox molecules can diffuse through the layer of DNA and interact with the diamond electrode to cause a redox current. By hybridization of DNA the intermolecular space shrinks which leads to Coulomb repulsion between negatively charge $Fe(CN)_6^{3-/4-}$ and negatively charged sugar-phosphate backbones of hybridized DNA. The redox reaction with diamond will therefore decrease (Fig. 45b).

The result is shown in Fig. 46 where we used cyclic voltammetry on H-terminated metallically doped (p-type) single crystalline diamond in 0.5 mM $Fe(CN)_6^{3-/4-}$, 100 mM KCL, 100 mM $KNO_3$ measured with respect to Ag/AgCl with a scan rate of 100 mV/s. The H-terminated diamond shows a well pronounced oxidation peak at +280 mV and a corresponding reduction wave with a peak at +126 mV (not shown here). These characteristics are well known and have been published in the literature [21-23]. After electrochemical attachment of phenyl linker molecules and ss-DNA marker molecules the redox amplitude is decreasing to about 30 % of the clean diamond surface. The peaks are not significantly shifted in potential or broadened by chemical modifications of the electrode (see Fig. 46). By hybridization, peaks are slightly shifted towards higher oxidation and lower reduction potentials. The change in amplitude is about 50 $\mu A/cm^2$. This change in voltammetric signal is reproducible and can be detected for several hybridization/denaturation cycles. In the future, this technique needs to be characterized in depth to evaluate the sensitivity, durability and reproducibility of this sensor array.

Yang et al. reported in 2004 [49] about cyclic voltammetry experiments using $Fe(CN)_6^{3-/4-}$ on boron-doped nanocrystalline diamond films coated with amines as linker and DNA. After ss-DNA marker attachment their redox currents decreased drastically and they conclude that the application of cyclic voltammetry is inhibited by the highly insulating nature of the molecular amine-layer linking DNA molecules to diamond. Our voltammetric experiments show that a detailed control of phenyl-molecule deposition is required. Insulating properties with respect to $Fe(CN)_6^{3-/4-}$ are detected if the phenyl-layer is grown slowly and thick, to form a dense scaffold on diamond. By short time attachment, using constant potential attachment technique, dispersed layers are generated so that diffusion of $Fe(CN)_6^{3-/4-}$ is not suppressed.

Alternatively, Yang et al. 2004 [49], Hamers et al. 2004 [81] and Gu et al. [50] applied impedance spectroscopy where they detect hybridization induced variations at low frequency. We have also applied such a detection scheme on our grafted diamond layers. A typical result is shown in Fig. 47. A clear difference between single-strand DNA bonded to diamond and double-strand DNA is detected. This technique allows discriminating hybridization of matched and mismatched DNA. The interpretation of these data requires, however, to know details about dielectric variations of ss-DNA and ds-DNA layers, about thickness variations by applied external electric fields [82], as well as the effect of redox-molecules like $Fe(CN_6)^{3-/4-}$ on the dielectric and conductivity properties of DNA films on diamond.

## 4. Bonding and detection of enzymes and proteins on diamond

After establishment of surface functionalization techniques for diamond using photo- and electrochemical means, bonding of molecules like enzymes [47] and proteins [48] has relatively fast been demonstrated. A detailed characterization of sensing properties of these devices is however still missing. As the use of carbon based transducers like glassy carbon, highly oriented pyrolytic graphite (HOPG), diamond like carbon and boron doped diamond is well established in electrochemistry, increasing biosensing applications of such materials can be expected. While the past has been dominated to find chemical treatments to achieve surface modifications of diamond, the future will be dominated by optimization of multi-layer structures for bio-sensing. Whether diamond will be the material of choice for bio-sensors will depend on the outcome of these research activities. The window of opportunities for diamond is now wide open, however, to compete with other transducer materials which are cheaper, less sophisticated in growth and compatible with silicon-technology will require optimized use of core advantages of diamond which are: a) chemical stability, b) strong bonding of organic molecules, c) low electrochemical background current, d) wide electrochemical potential window, e) no gate-insulator for DNA-FETs from diamond, and f) perfect surface electronic properties which includes control of interfacial energy alignments by H- or O-termination to optimize electron transfer reactions.

## 5.      Summery and conclusions

In this review, we have summarized recently achieved results with respect to photo- and electrochemical surface modifications of diamond. Both techniques have been optimized to a level, which allowed the realization of first generation electrochemical and field-effect DNA sensors. By AFM experiments we detect a formidable bonding stability of bio-molecular arrangements on diamond. It confirms earlier reports by Yang et al. 2002 [14] about exceptional DNA bonding stability in hybridization cycles. Applications of diamond bio-sensors in high-throughput systems where especially high bonding stability is required will therefore be of significant interest in the future. Our experiments also show that basic understanding of growth mechanisms, of electronic and chemical properties of each layer of the composite bio-recognition film (for example: amine/crosslinker/ss-DNA) is required to achieve progress with respect to optimization of sensor performance.

Finally, the proper selection of "diamond transducer materials" for bio-sensor applications is of comparable importance. Electronic detection of bonding events in electrolyte solutions requires high quality diamond with minimized defect densities, no grain-boundaries and no sp2, and with a defect free surface termination. Such diamond transducers are and will be more expensive than established semiconducting materials like Si. We assume therefore, that future applications of diamond will be in clinical, high through-put systems which require high bonding stability. Multi-array sensors from single crystalline diamond can be realized using established technologies and chemistry. As prices for established DNA multi-array optical sensors are rather expensive (typically > 5000 Euro, see http://www.affymetrix.com), the costs for diamond substrates of typically 150 to 250 Euro will not be a strong argument against the use of diamond for such applications. In any case, diamond needs to show that its device properties and performances are superior compared to other transducer layers to be commercially successful.

There are several fields where nano- and polycrystalline diamond seems to become a promising leading application. Nano-size diamond particles with a diameter of typically 10 nm are currently investigated as core material for rapid, low volume solid phase extraction of analytes, including proteins and DNA from a variety of biological samples [95-100]. These particles are unique as they are optically transparent, carry specific functional groups and can mix rapidly with sample solutions upon agitation. After extraction of the target analyte by these particles, the diamond nano-particles can be recollected and analyzed. Second, nano-diamond is capable to be used as color-center marker as it has several interesting defect related light emitting centers ("color centers"), like the nitrogen-vacancy (N-V) complex and others, which makes nano-diamond unique with respect to fluorescence microscopy applications. These color-centers are resistant against photo-bleaching and can be up-taken by mammalian cells with minimal cytotoxicity [95,100].

The increasing demand for secure, mobile, wireless communication has stimulated interest in technologies capable of reducing the size and power consumption of wireless modules, and enhancing the bandwidth efficiency of communication networks. Nano-diamond micro-electrical-mechanical-systems (MEMS) are at the leading edge of this field as the frequency (f) quality (Q) product of such nano-diamond oscillators reached f = 1.51 GHz and the quality factor Q = 11,555 [101-103] In the field of bio-sensing, such high quality mechanical oscillation systems are also of significant importance for improved detection and sensibility [104,105].

Direct manipulation of living cells or transfer of molecules into cells ("cell surgery") is an emerging and increasingly important technology in biology. It requires new tools with dimensions below the micrometer regime. A typical gene surgery tip should be about 40 μm long with a diameter of around 400 nm (for a review see Ref. 106). In addition, the material should not poison the cell during manipulation, the surface should be optimized with respect to friction of cell membranes and it should allow applying given potentials to the tip to electrostatic bond or releasing DNA fragments. Best candidates for these applications are nano- and poly-crystalline diamond as they show i) the required mechanical stability (hardness 50 100 GPa), ii) the surface can be adjusted to optimize friction and iii) if properly designed, the core of the tip can be conductive by boron-doping.

Based on these arguments, we are convinced that bio-application of diamond either monocrystalline for electronic sensing or poly- and nano-crystalline for mechanical techniques, are very realistic opportunities for this promising material. Diamond will surely find its place in the rapidly growing field of biology and bio-technology.

*Acknowledgement*

The authors want to thank T. Nakamura for the synthesis of TFFAD molecules, which contributed significantly our progress as well as to T. Yamamoto, who helped to fabricate DNA-FET detectors. We also want to thank Dr. Park, who supported our activities with an electrochemical analyzer system, Dr Watanabe, Dr. Ri and Dr Tokuda for growth of excellent diamond films. The authors thank The Journal of the Royal Society Interface for partially copyright transfer of our review paper with the title: diamond and biology, published ASPA (http://unit.aist.go.jp/dia-rc/dia-surf/index.htm) and .

**Figure Captions**

**Fig. 1**

Voltammograms for water electrolysis on various electrodes. The supporting electrolyte is 0.5 M $H_2SO_4$. The graphs are shifted vertically for comparison. Two poly-crystalline films, B:PCD(NRL) with $5x10^{19}$ $B/cm^3$ and B:PCD(USU) with $5x10^{20}$ $B/cm^3$ (from Ref. 21 and 22) are compared with a single crystalline boron-doped diamond B:(H)SCD with $3x10^{20}$ $B/cm^3$ and with an undoped diamond (H)SCD. Also shown are data for Pt, Au and glassy carbon from Ref. 18. Oxidation reactions, e.g., oxygen evolution, have positive currents and emerge around 1.8 V for all diamond samples. Reduction reactions, e.g., hydrogen evolution, have negative currents and show very different properties. Note, the background current within the regime between hydrogen and oxygen evolution for diamond is very low, and the electrochemical potential window large, compared to glassy carbon, Pt and Au.

**Fig. 2**

Energies of the conduction- and valence-band edges of a number of conventional semiconductors, and of hydrogen terminated and hydrogen-free diamond relative to the vacuum level $E_{VAC}$ are shown. Please note, H-terminated diamond shows a negative electron affinity $(-\chi)$ as the conduction band edge is above the vacuum level of the electrolyte. The dashed horizontal line marks the chemical potential $\mu$ for electrons in an acidic electrolyte under conditions of the standard hydrogen electrode. The insert shows the chemical potential under general non-standard conditions as a function of pH and for different partial pressure of hydrogen in the atmosphere as given by Nernst's equation [29].

**Fig. 3**

a) Fermi-level and chemical potential alignment at the interface diamond/electrolyte after equilibration. Due to transfer doping, electrons are missing in diamond so that a thin two-dimensional (2D) hole accumulation layer is generated [32,33]. The hole density in this layer depends on the chemical potential as indicated by arrows.

b) pH-sensitivity of a diamond ion-sensitive field effect transistor (ISFET) [34,35,39-41]. The gate potential shift shows a pH-dependence of 55 mV/pH, which is close to the Nernst prediction.

**Fig. 4**

Stability of DNA bonding to ultrananocrystalline diamond, Au, Si and glassy carbon as detected during 30 successive cycles of hybridization and denaturation. In each case the substrates were amine-modified and then linked to thiol-terminated DNA [14].

**Fig. 5**

Comparison of different diamond films: a) Ultranano-, nano- and b) polycrystalline diamond layers are dominated by grain-boundaries which are decorated with sp2 and amorphous carbon [54,56-58]. The volume-fraction of sp2 and grain boundaries depends on growth parameter and varies from layer to layer. Amorphous carbon and sp2 generate a continuous electronic density-of-states distribution in the gap of diamond. In addition, such diamond films show a significant surface roughness in the rang of 30 to 50 nm for nano-crystalline diamond (see a, II) and micrometer to tens of micrometer for polycrystalline layers (b, II). On the other hand, single crystalline CVD diamond has been optimized over recent years to electronic grade quality with atomically smooth surfaces (c, II) [63-65]. A typical cathodoluminescence spectrum measured at 16 K is shown in c, I.

**Fig. 6**

Typical AFM surface morphology of a single crystalline CVD diamond surface, as detected by AFM and used in these studies. The root mean square (RMS) roughness below 1 Å.

**Fig. 7**

Typical temperature dependent conductivity of a metallically boron doped CVD diamond. The doping level is in the range of $5 \times 10^{20}$ B/cm$^3$. σ is activated with 2 meV which indicates hopping propagation of holes in the acceptor band.

**Fig. 8:**

Typical geometry and arrangement of realized CVD diamond DNA ions-sensitive field-effect-transistor (FET) structures (DNA-ISFET). The image has been generated by scanning electron microscopy (SEM), where the sensor area ("gate") is H-terminated and surrounded by oxidized diamond. Drain and source contacts are from Au. The H-terminated area is photochemically modified to bond ss-DNA marker molecules covalently to diamond.

**Fig. 9**

10-amino-dec-1-ene molecule protected with trifluoroacetic acid group ("TFAAD"), as determined by Molecular Orbital Calculations [74].

**Fig. 10**

Front (a) and side view (b) of a nitrophenyl molecule, as calculated by Molecular Orbital Calculations [74].

**Fig. 11**

Electrochemical (I) and photochemical (II) boning schemes:

Scheme I: a) Nitrophenyl linker molecules are electrochemically bonded to H- or O-terminated diamond. Nitrophenyl is reduced to aminophenyl and reacted with a hetero-bifunctional crosslinker. Finally thiol-modified ss-DNA is attached.

Scheme II: Amine molecules are photochemically covalently attached on H-terminated diamond. The linker molecules are then deprotected and reacted with hetero-bifunctional crosslinker molecules and thiol-modified ss-DNA.

**Fig. 12**

DNA-FET sample holder arrangement. The diamond sensor is mounted in a PEEK plate as shown in a). The drain and source pads are contacted using Pt wires. A second PEEK part (b) with silicon rubber is mounted on to of part a) and closed to seal the drain source pads from electrolyte buffer c). A Pt wire is used as gate electrode and the exposed sensor area is 0.7 mm x 1mm.

**Fig. 13**

AFM topography and line scan of a TFAAD film attached by 1.5 hours of illumination to diamond. The scratched area is $2\times2\mu m^2$ and the loading force was 200 nN.

**Fig. 14**.

Thickness of TFAAD films, attached photochemically to H-terminated single crystalline diamond, as a function of illumination time. Each averaged thickness was measured at four different scratched areas. Please note, that the indicated bars represent the width of height variations.

**Fig. 15**

a) XPS survey spectrum of a hydrogen-terminated single crystalline diamond surface that was exposed to TFAAD and 20 mW/cm$^2$ UV illumination (250 nm) for 2 hours. b) The C(1s) spectrum reveals two additional small peaks at 292.9 eV and 288.5 eV which are attributed to carbon atoms in the CF$_3$ cap group and in the C=O group, respectively. c) The ratio of the F(1s) signal (peak area) to that of the total C(1s) signal as a function illumination time is time dependent, follows an exponential increase (dashed line), with a characteristic time constant $\tau$ of 1.7 h. d) Angle resolved (with respect to the surface normal) XPS experiments show an increase of the F(1s)/C(1s) peak intensities, rising from 48º to 78º.

**Fig. 16**

Variation of TFAAD film thickness as detected by AFM (circles) and the variation of the integrated peak intensity ratio F(1s)/C(1s) (squares) from XPS experiments as a function of illumination time. Triangles show the onset of cross-polymerization of TFAAD molecules.

**Fig. 17**

Side and top view of a reconstructed diamond (100)(2x1):1H surface (data from Ref. 78). Also shown are typical areas of TFAAD molecules, assuming a diameter of 5.04 Å as shown in Fig. 9.

**Fig. 18**

Comparison of diamond ion-sensitive field effect transistor properties (ISFET) measured in SSPE buffer with perfectly H-termination of the surface (circles) and after photo-attachment of amine molecules for 20h (squares). The drain-source current is decreasing for about 60 %, but the surface of diamond remains conductive after photo-attachment.

**Fig. 19**

Schematic diagram of the photoexcitation mechanism at the surface of diamond in contact with TFAAD (from refs. 70,71). Valence-band electrons are photoexcited into empty surface states of diamond and then into empty electronic states of TFAAD molecules, which generates nucelophilic properties.

**Fig. 20**

Schematic grafting mechanism of the diamond surface by amines (from refs. 77, 80). i) shows the generation of radical anions by electron transfer from diamond to the olefin (a). The nucleophilic properties of the radical causes a hydrogen abstraction (b) which results in a surface carbon dangling bond; ii) The dangling bond reacts with an olefin molecule (c) to form a diamond-carbon/olefin-carbon bond. This olefin abstracts a hydrogen atom from the diamond surface (d) which is a new site for olefin addition; iii) The hydrogen abstraction reaction results in a chain reaction; iv) In case of extended illumination (> 10 h) a 3D growth sets in due to cross-polymerization of olefin molecules.

**Fig. 21**

Fluorescence microscopy image of double stranded (ds) DNA helices on diamond using green fluorescence tags (Cy5) attached to complementary DNA. The bright areas arise from initially H-terminated diamond, the less intense regions were originally oxidized. Black areas are Au contacts.

**Fig. 22**

Typical surface morphology of a DNA film on diamond after application of contact mode AFM (a), where increasing AFM tip loading forces have been applied. Forces larger than 45 nN gives rise to DNA removal on photochemically treated and initially H-terminated diamond. After removal of DNA from the surface it appears dark in fluorescence microscopy (b).

**Fig. 23**

Oscillatory AFM measurement applied at the boundary of cleaned diamond surface to initially oxidized (left) and initially H-terminated (right) diamond show that on both areas DNA molecules are present. The height on O-terminated diamond is however lower than on H-terminated diamond. Molecules on oxidized diamond can be removed with forces of about 5 nN.

**Fig. 24**

(a) Optimized oscillatory AFM measurement at the boundary of a cleaned diamond surface to diamond with attached double-stranded DNA molecules. The squares denote the regions where AFM phase shifts were evaluated. (b) AFM height profile across the boundary reveals a DNA layer thickness of 76 Å (see also Ref. 75).

**Fig. 25**

AFM set-point ratio dependence of AFM measurements of DNA height and phase contrasts across a DNA-functionalized and cleaned diamond surface for free oscillation amplitudes ($A_o$) of 6 and 10 nm. Extrapolation to set-point ratio 1 results in a DNA height of about 76 Å.

**Fig. 26**

From AFM, a compact DNA layer of 76 Å height is resolved by optimizing phase and height contrast measurements. The axis of the double helices DNA is therefore tilted by about 30 to 36° with respect to the diamond surface as shown schematically in this figure.

**Fig. 27**

AFM height profile shows a dense DNA layer with a RMS height modulations of $\pm$ 5 Å.

**Fig. 28**

Variation of the ss-DNA marker molecule attachment time, between 10 minutes and 12 h gives rise to an exponential increasing fluorescence intensity. The fluorescence intensity of hybridized DNA follows an activated property (full line) with a time constant of 2 h. The DNA density on diamond varies between $10^{12}$ and $10^{13}$ cm$^{-2}$.

**Fig. 29**

Cyclic voltammograms for 1 mM 4-nitrobenzene diazonium tetrafluoroborate on highly B-doped single-crystalline diamond (DRC). Electrolyte solution: 0.1 M NBu$_4$BF$_4$ in CH$_3$CN. scan rate: 0.2 V/s (for details see Ref. 72).

**Fig. 30**

Comparison of cyclic voltammorgams for 1 mM 4-nitrobenzene diazonium tetrafluoroborate measured on H-terminated (red) and oxidized (green) highly B-doped single-crystalline diamond. Please note, that the attachment of phenyl molecules to oxidized occurs at slightly more negative potentials of about -410 mV compared to H-terminated diamond at -170 mV (vs. Ag/AgCl).

**Fig. 31**

Cyclic voltammograms from 4-nitrophenyl modified single-crystalline diamond in blank electrolyte solution. Electrolyte solution: 0.1 M NBu$_4$BF$_4$ in CH$_3$CN. For details see Ref. 72.

**Fig. 32**

Schematic reduction/oxidation reactions of nitrophenyl bonded to diamond, giving rise to a two electron transfer reaction mechanism [72].

**Fig. 33**

Transient current as detected during nitrophenyl attachment at a constant potential of -0.2 V (vs. Ag/AgCl). Also shown is the theoretical decay following a $t^{-0.5}$ time dependence. The inset shows a schematic growth model where the phenyl layer starts to grow (1), to become thicker (2) and finally terminates the growth (3), as the electron tunneling transition through the phenyl layer decreases to zero [76].

**Fig. 34**

AFM scratching experiments on nitrophenyl modified diamond. With forces > 100 nN most of the phenyl-layer can be removed while forces > 120 nN are required to remove the linker layer to diamond completely [76].

**Fig. 35**

Nitrophenyl layer growth during constant potential attachment experiments with -0.2 V (vs. Ag/AgCl) is governed by three dimensional (3D) growth properties. After short time attachment the layer thickness varies strongly, whereas after 90 s the variations become much smaller, indicating a dense layer formation of about 25 Å thickness (for details see [76]).

**Fig. 36**

Angle resolved XPS experiments show oriented growth of nitrophenyl-layers, grown with constant potential. Cyclic potential attachment method gives rise to significantly thicker layers of typically 30 to 70 Å with less pronounced molecule arrangement.

**Fig. 37**

Nitrophenyl groups at an initial stage of attachment grow three-dimensional (3D) as shown here schematically, forming layers of varying heights and densities. Layer thicknesses of up to 80 Å are detected for cyclic voltammetry attachment after 5 cycles, whereas the layer becomes denser and only about 25 Å thick in case of constant potential attachment.

**Fig. 38**

The nitro-groups ($NO_2$) are electrochemically changed to amino-groups ($NH_2$) by cyclic voltammograms in 0.1 M KCl solution with 10:90 (v/v) $EtOH-H_2O$. Scan rate:

0.1 V/s. During the first sweep, a pronounced peak is detected which reflects the reduction of $NO_2$ to $NH_2$. The variation in the second and third sweep is minor compared to the first sweep reduction.

**Fig. 39**

Fluorescence microscopy image of a ds-DNA-functionalized single crystalline diamond electrode after DNA hybridization with complementary oligonucleotides terminated with Cy5 dye molecules. The layer was originally H-terminated but a T-shape has been oxidized. Here, the fluorescence intensity appears weaker. To generate a contrast with respect to clean diamond, a scratched area has been realized.

**Fig. 40**

Comparison of critical removal forces of electrochemically attached ds-DNA on H-terminated and oxidized diamond and of photochemical attached ds-DNA on H-terminated and oxidized diamond. Attachment to initially H-terminated diamond of both linker molecules systems gives rise to stronger bonding than to initially oxidized diamond surfaces. DNA bonded with phenyl linker molecules show the strongest bonds.

**Fig. 41:**

Comparison of DNA removal forces as detected in our experiments on diamond and compared with DNA bonding to Au (a: [83], b: [84], c: [85]) and Mica (d: [86]).

**Fig. 42**

Schematic description of DNA hybridized on a diamond DNA ion-sensitive field-effect-transistor (DNA-ISFET) sensor. The surface conductivity of diamond will be change by accumulation of compensating cations in the DNA layer which is caused by the negatively charged back-bone structure of DNA (for details see ref. 84). The 1M NaCl buffer shrinks the Debye length in our experiments to about 3 Å.

**Fig. 43**

a) Drain-Source current variations measured as a function of gate potential at a fixed drain-source potential of -0.5 V for ss-DNA (marker-DNA), after hybridization with complementary target ss-DNA to form ds-DNA and after removal of DNA by washing. A gate potential shift of about 80 mV is detected on this DNA-ISFET with about $4 \times 10^{12}$ cm$^{-2}$ molecules bonded to the gate.

b) Gate-potential shifts as detected on diamond transistor structures with $10^{12}$ cm$^{-2}$, $4 \times 10^{12}$ cm$^{-2}$ and $10^{13}$ cm$^{-2}$ ss-DNA marker molecules bonded to the gate-area. The threshold potential is increasing towards less dense grafted diamond gates areas.

**Fig. 44**

Comparison of silicon based DNA-ISFET sensitivities as a function of data publication (black half filled squares from [79], and half-filled triangles from [87]) with diamond sensitivities as shown in this paper (half-filled diamonds, DRC) and with data deduced on polycrystalline CVD diamond films from ref. 88 (half-filled circle). The shaded area indicates the theoretically predicted sensitivity, following the model of Poghossian et al. 2005 [84].

**Fig. 45**

Schematic DNA hybridization detection mechanism using $Fe(CN_6)^{3-/4-}$ as mediator redox molecules. In case of ss-DNA (a), the negatively charged redox molecules $Fe(CN_6)^{3-/4-}$ (blue balls) can diffuse through the DNA layer. After hybridization (b), the space between individual ds-DNA molecules becomes to small for negatively charge molecules to overcome the repulsive Coulomb forces from the negatively charge back-bone of ds-DNA.

**Fig. 46**

Cyclic voltammograms on ss- and ds-DNA grafted metallically boron-doped (p-type) single crystalline diamond in 0.5 mM $Fe(CN_6)^{3-/4-}$, 100 mM KCL, 100 mM KNO$_3$ measured with respect to Ag/AgCl with a scan rate of 100 mV/s. The oxidation and reduction peaks are decreasing for about 50 μA/cm$^2$ by hybridization.

**Fig. 47**

Impedance spectroscopic properties of DNA-modified nano-diamond films. The impedance is shown in the complex plane as detected for ss-DNA, for exposure to 4-base mismatched DNA and after exposure to complementary DNA in pH 7.4 phosphate buffer containing 1mM $Fe(CN_6)^{3-/4-}$.

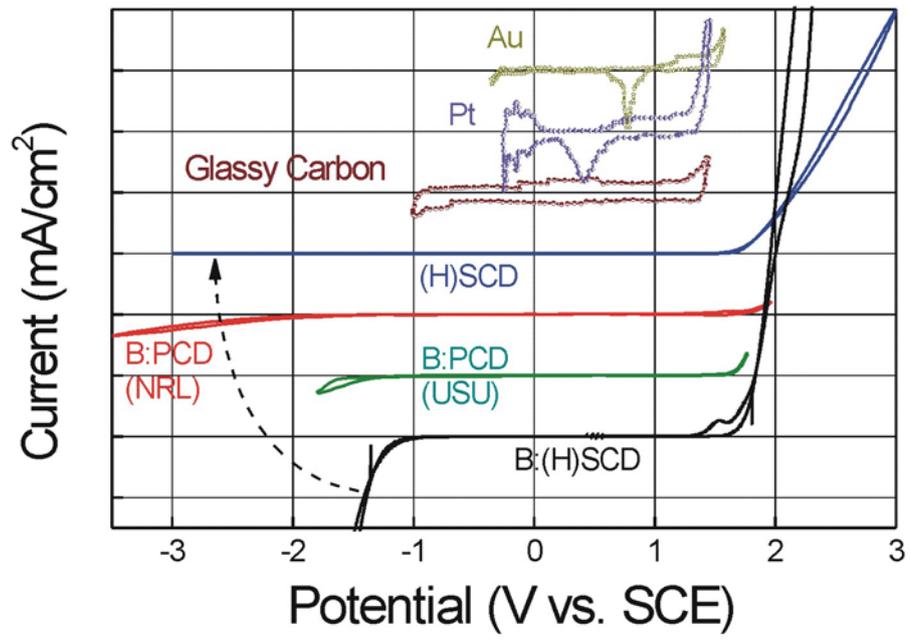

Fig. 1

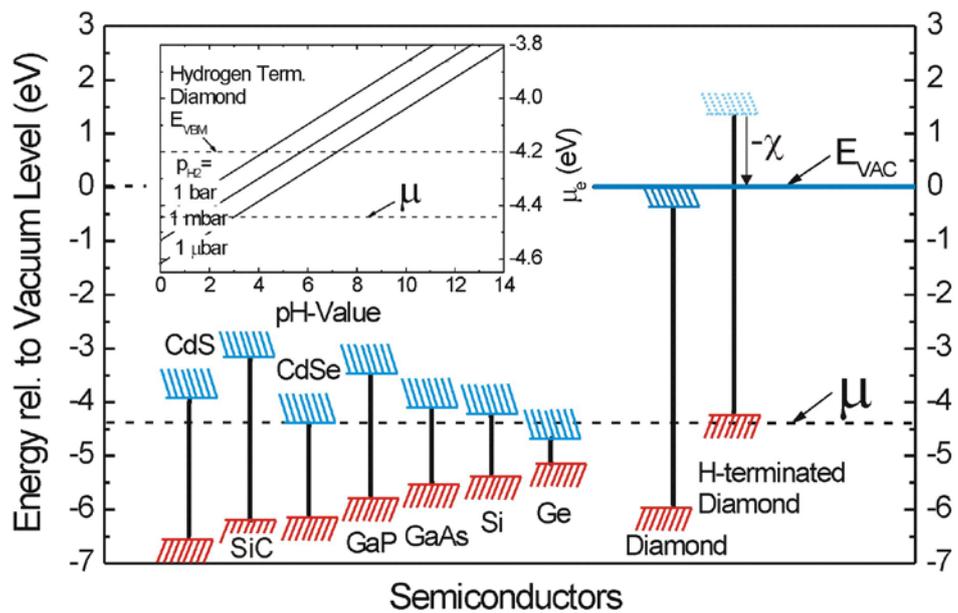

Fig. 2

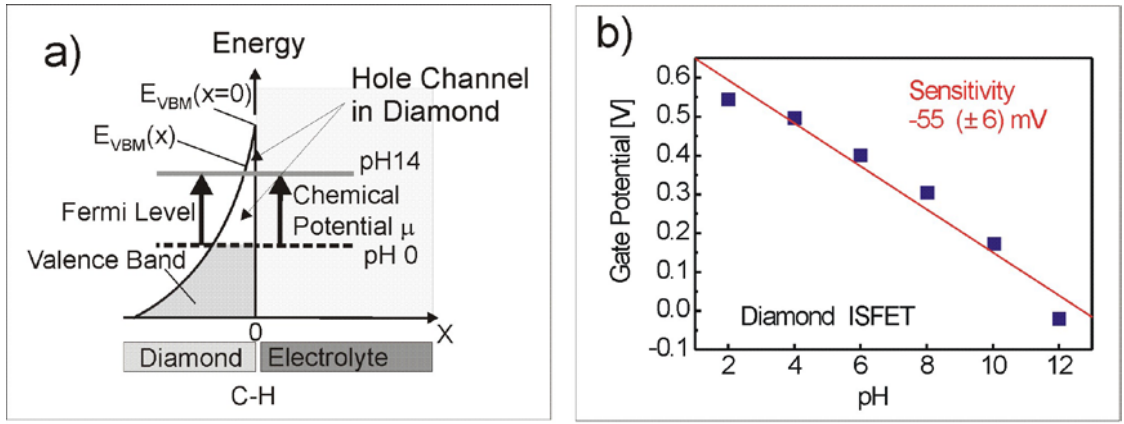

Fig. 3

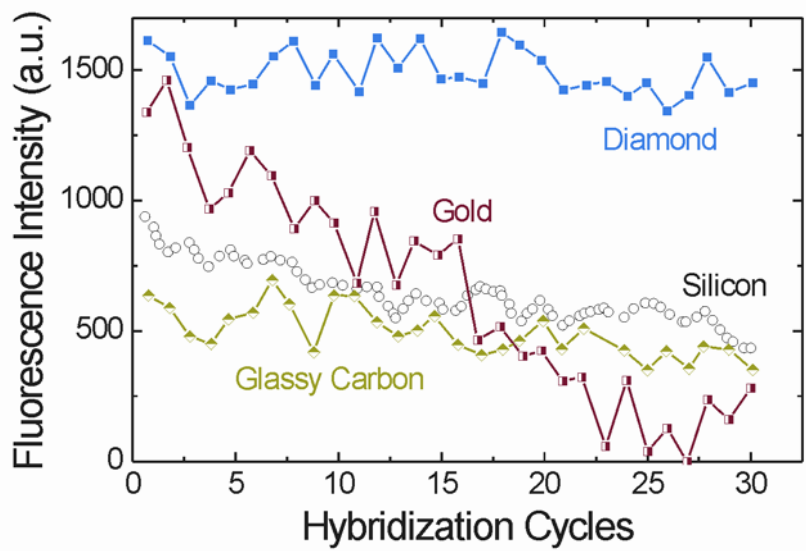

Fig. 4

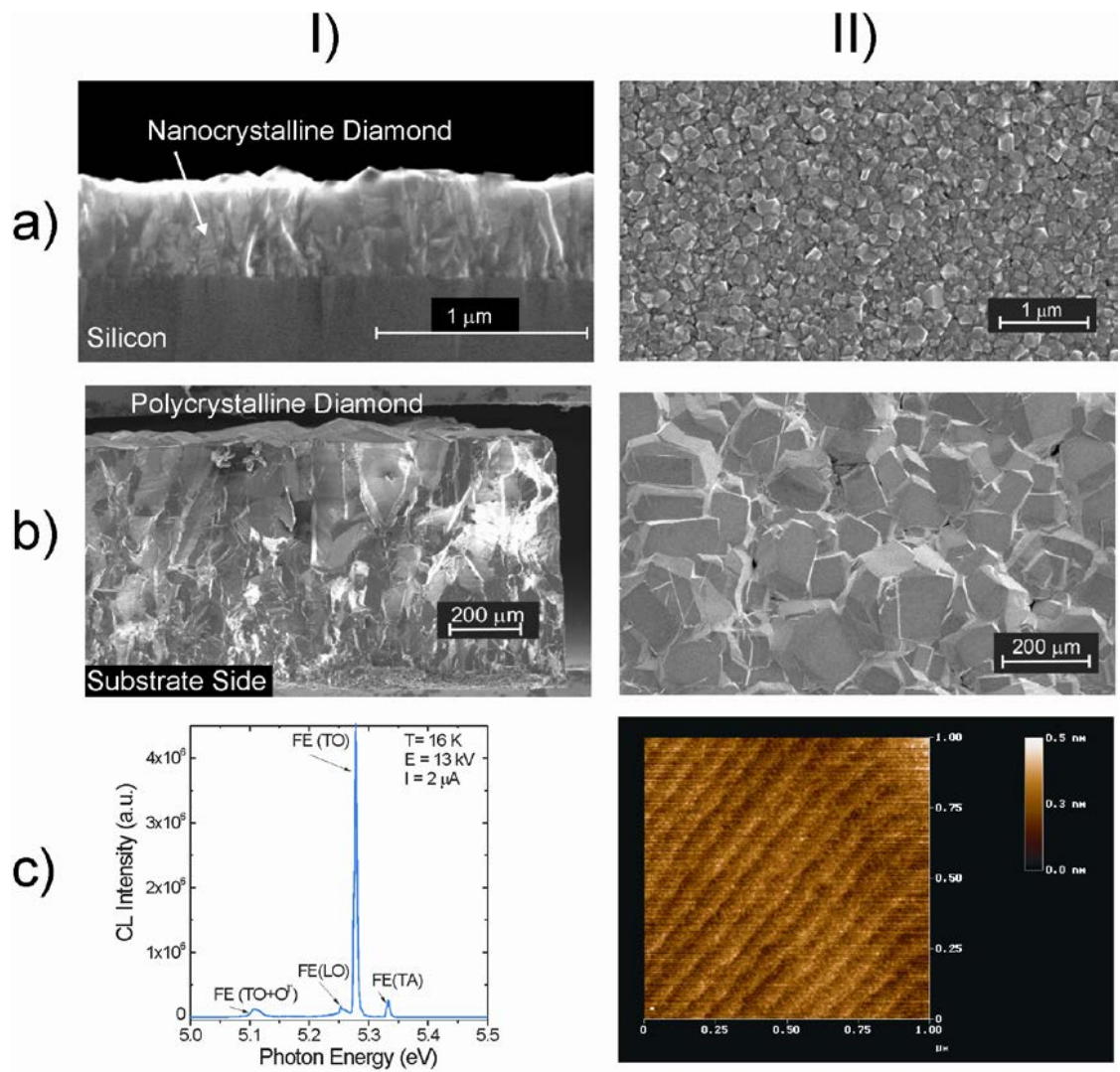

Fig. 5

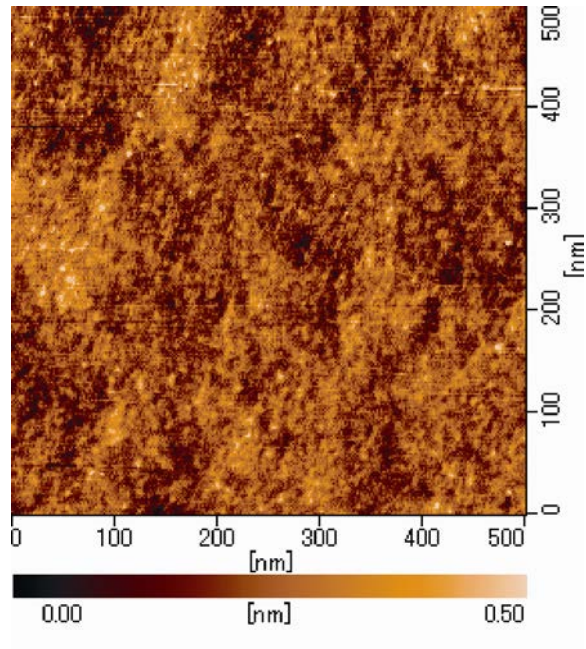

Fig. 6

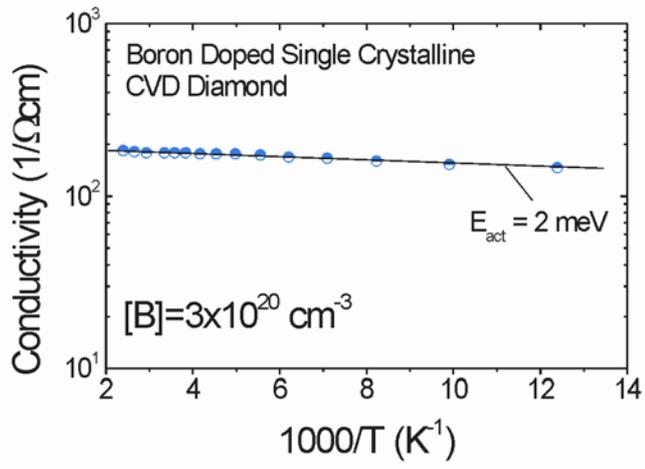

Fig. 7

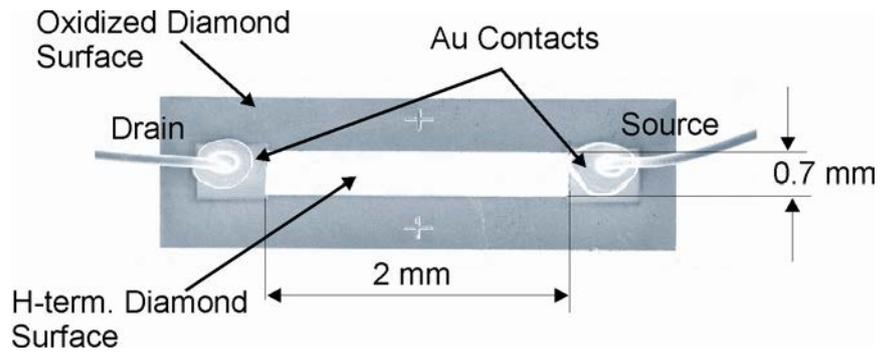

Fig 8

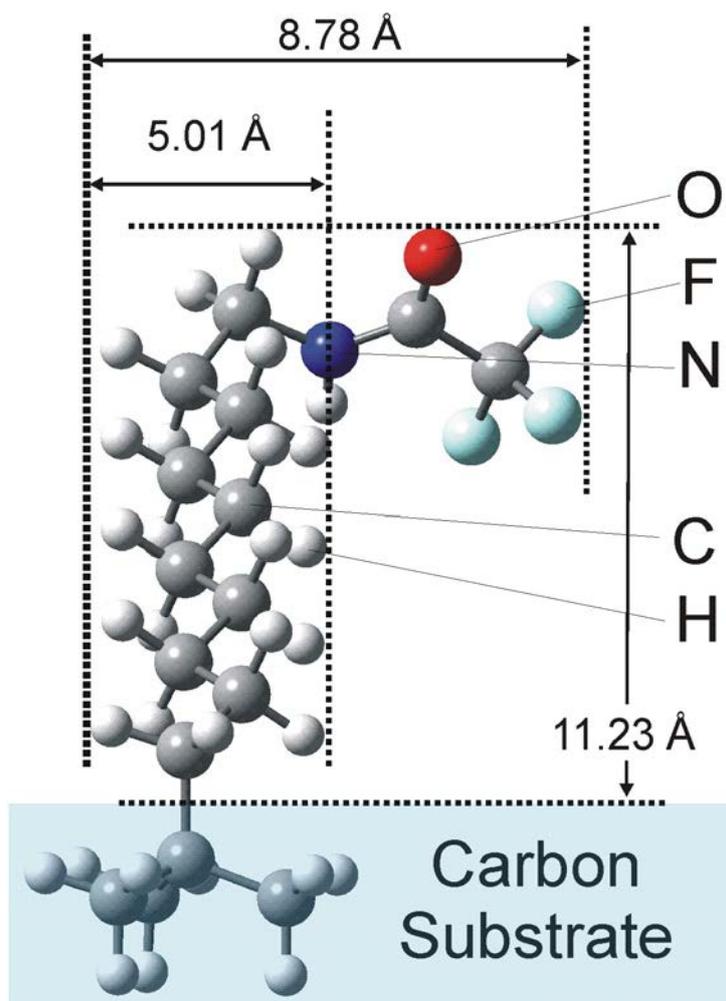

Fig. 9

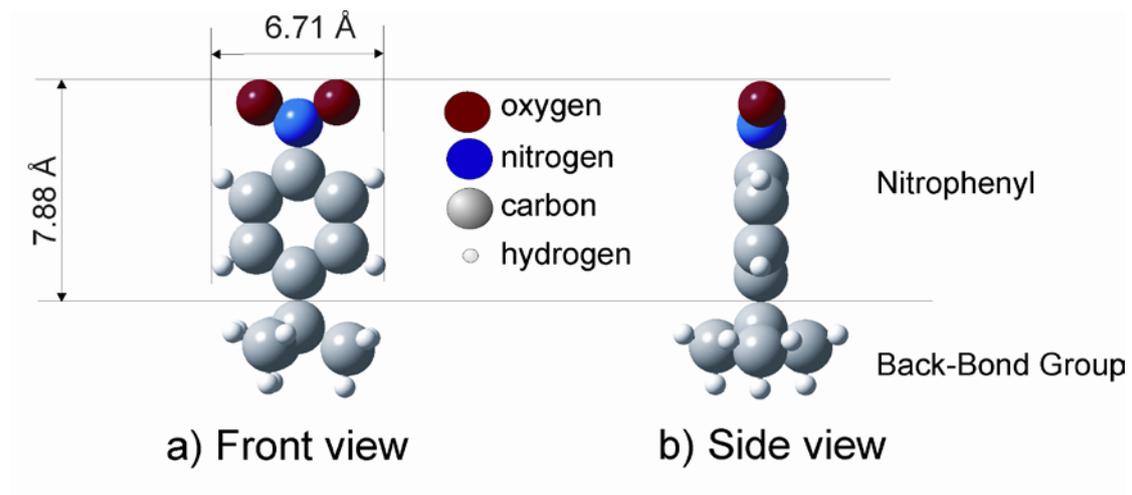

a) Front view        b) Side view

Fig. 10

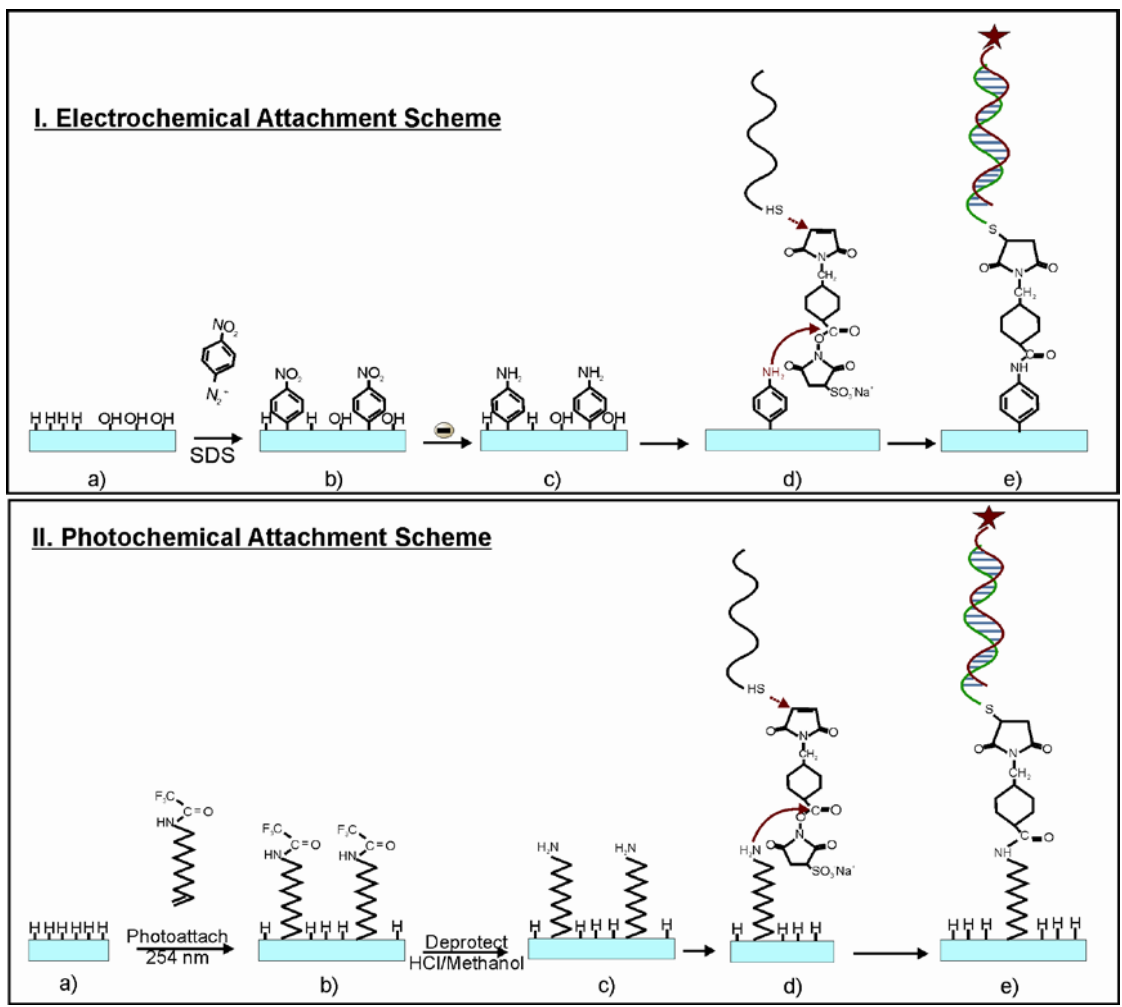

Fig. 11

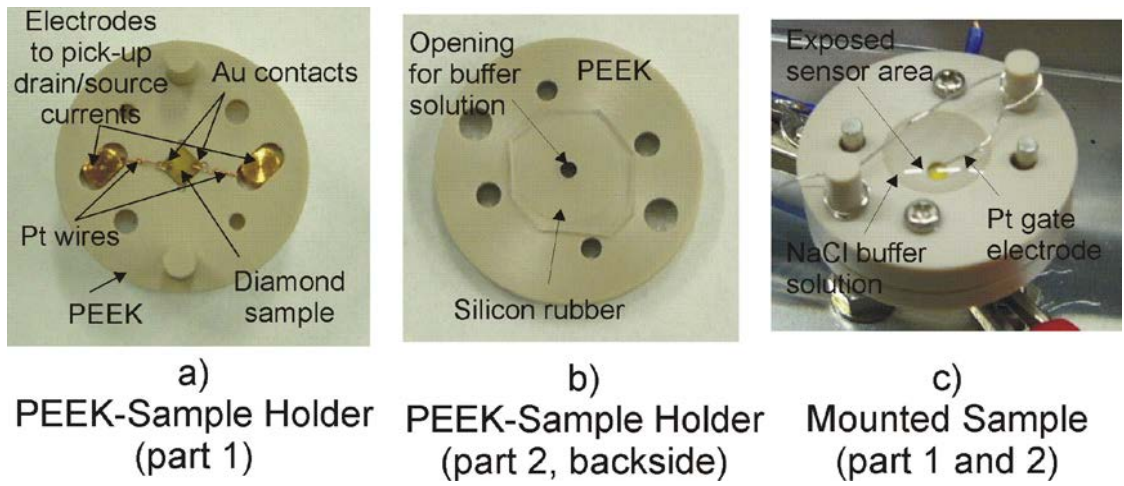

a)
PEEK-Sample Holder
(part 1)

b)
PEEK-Sample Holder
(part 2, backside)

c)
Mounted Sample
(part 1 and 2)

Fig. 12

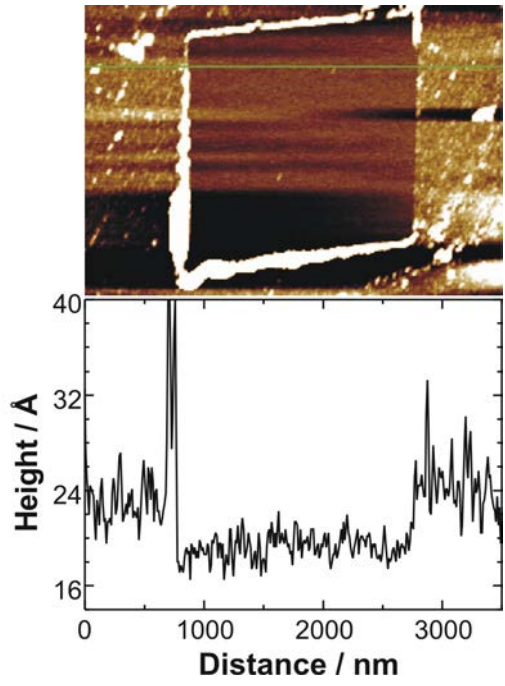

Fig. 13

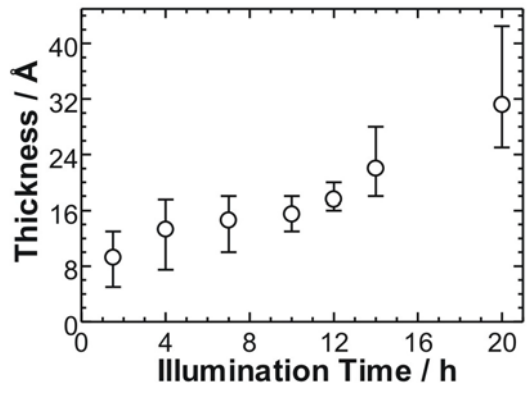

Fig. 14

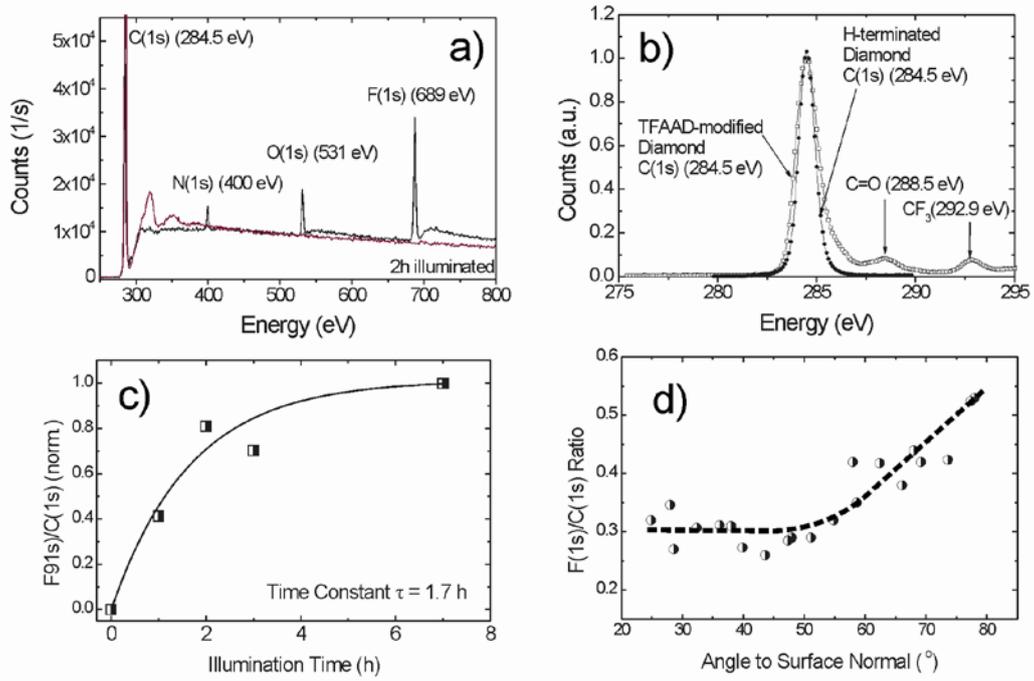

Fig. 15

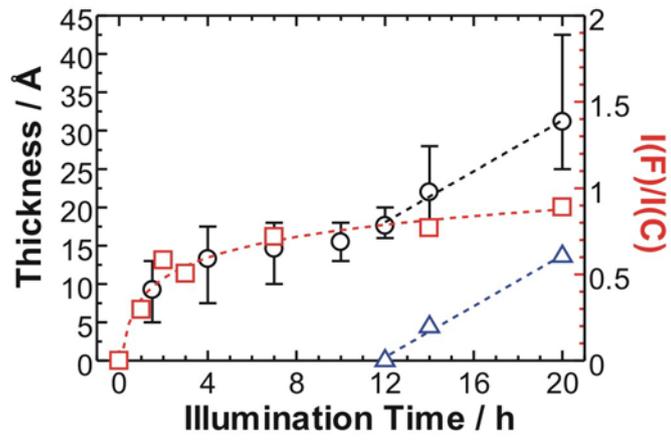

Fig. 16

## C(100)(2x1):1H

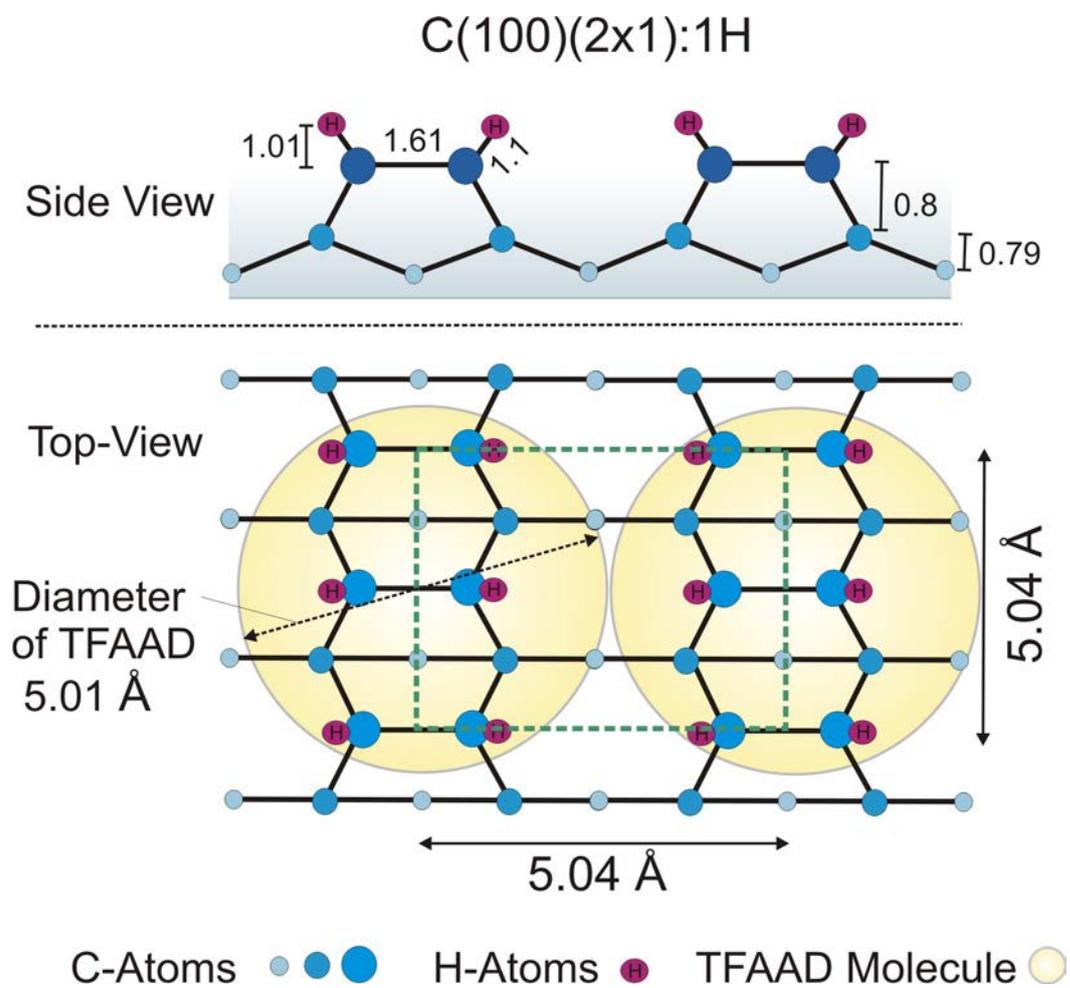

Fig. 17

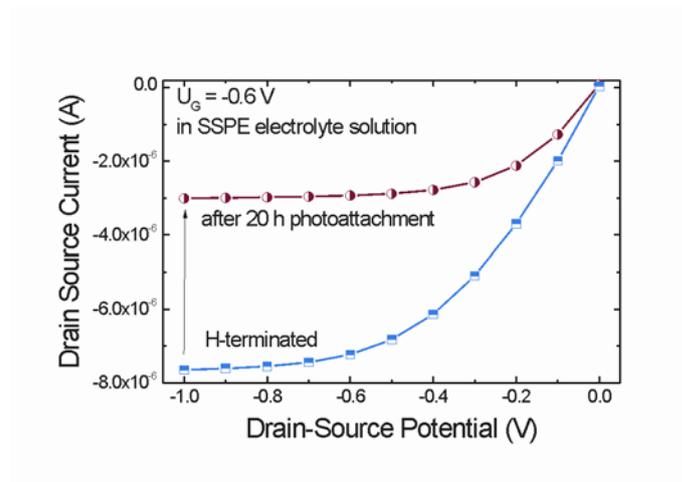

Fig. 18

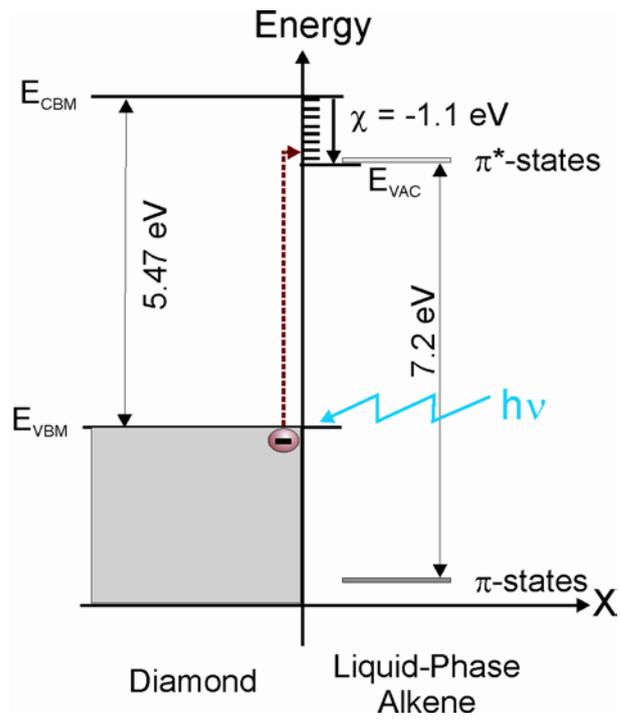

Fig. 19

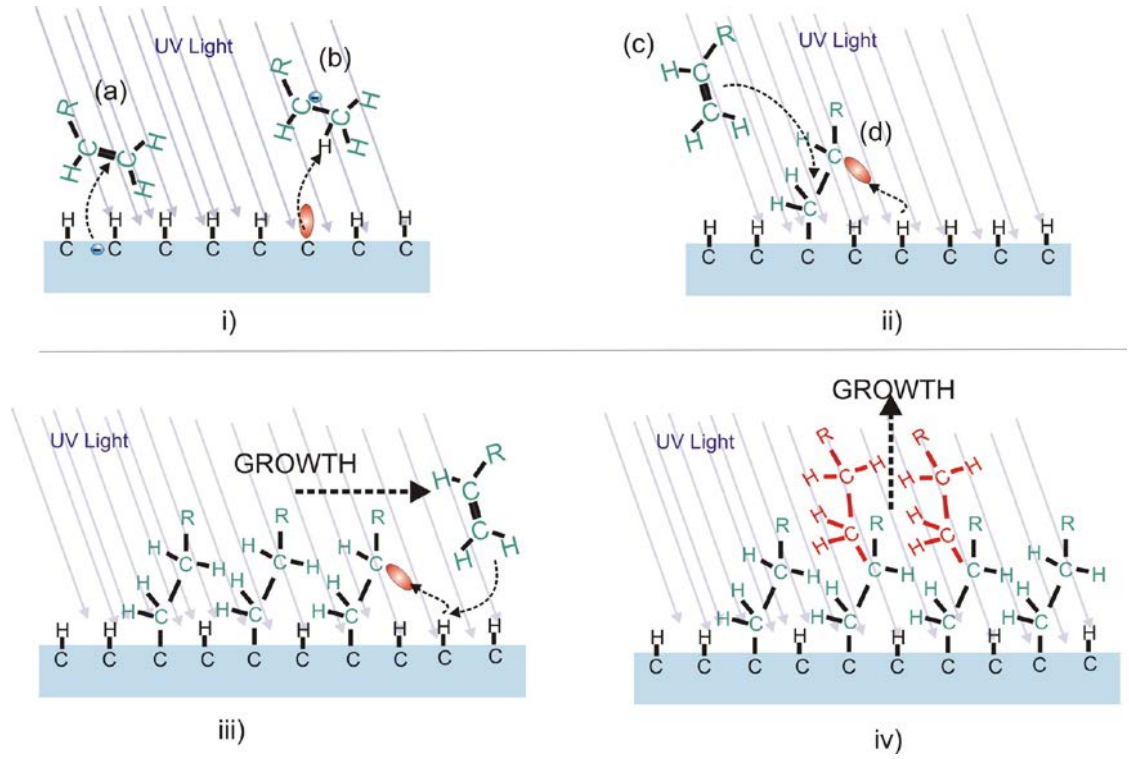

Fig. 20

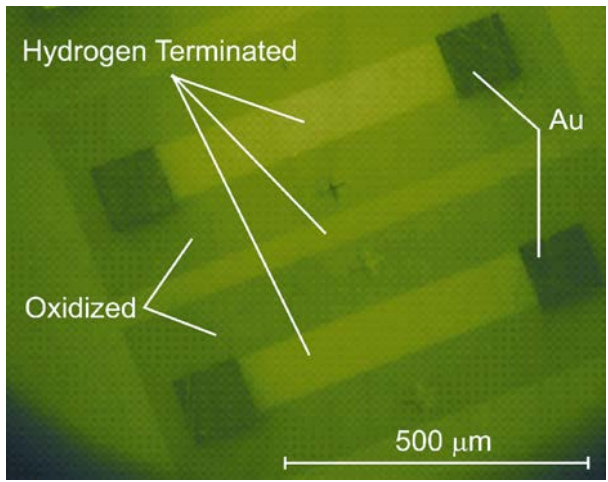

Fig. 21

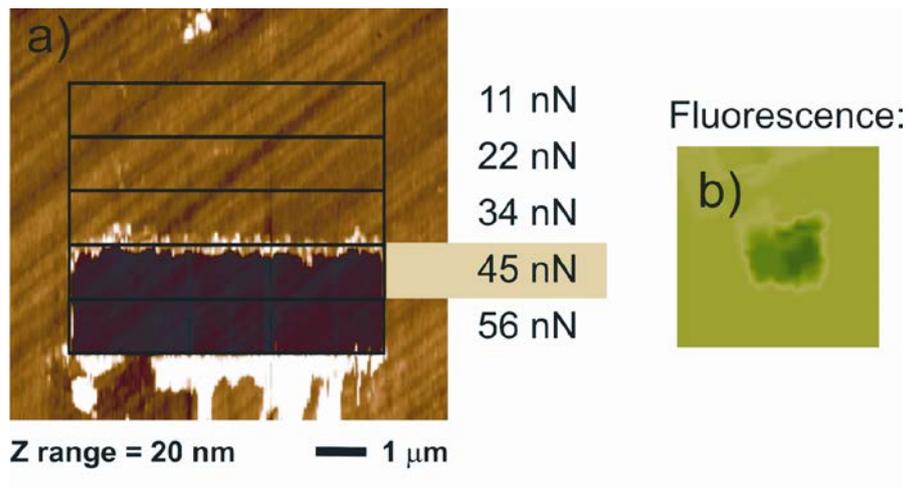

Fig. 22

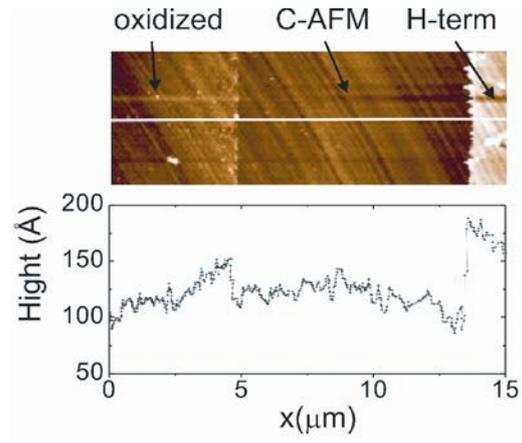

Fig. 23

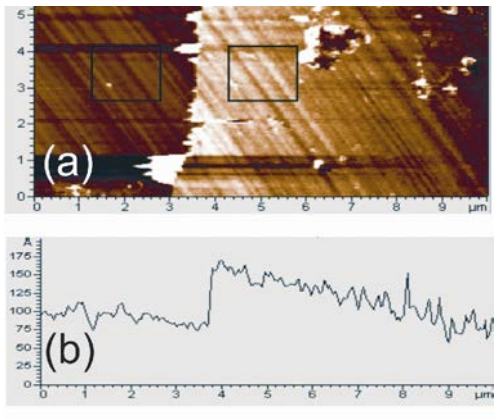

Fig. 24

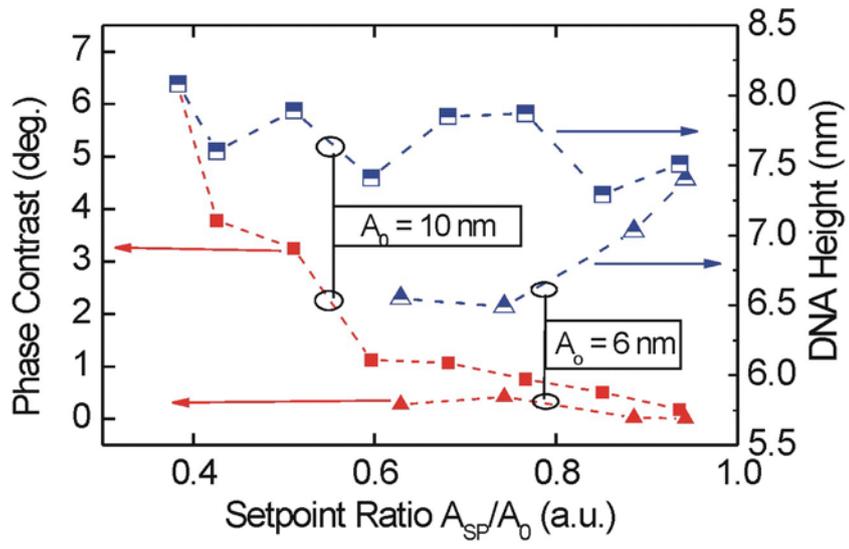

Fig. 25

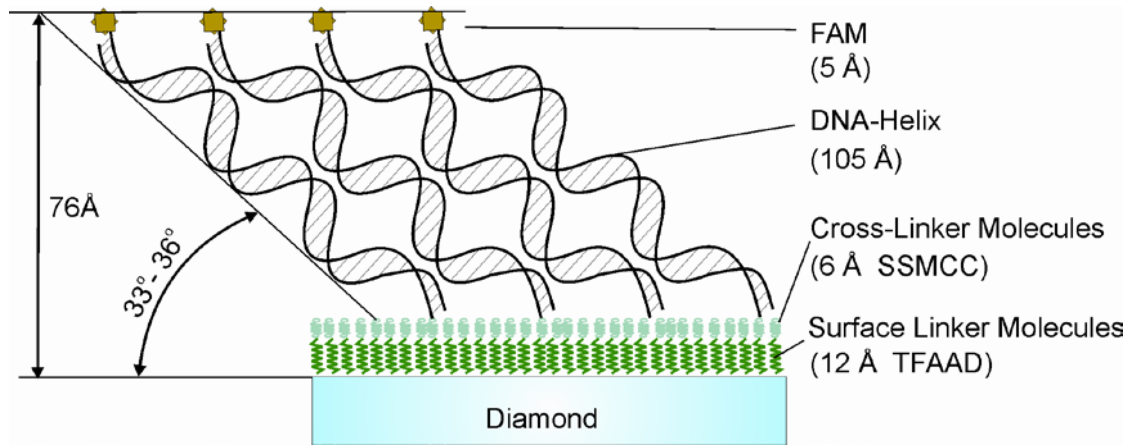

Fig. 26

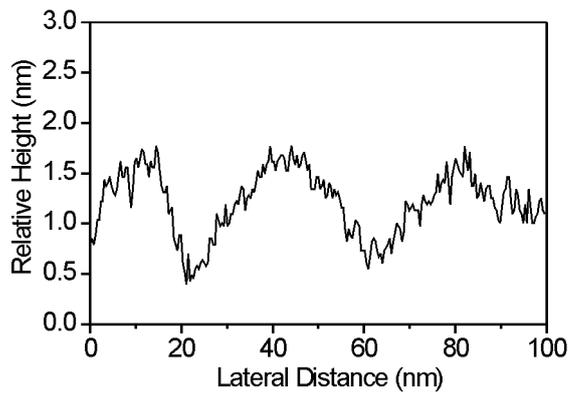

Fig. 27

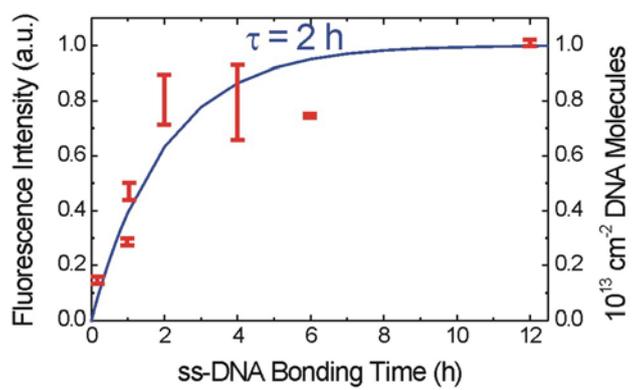

Fig. 28

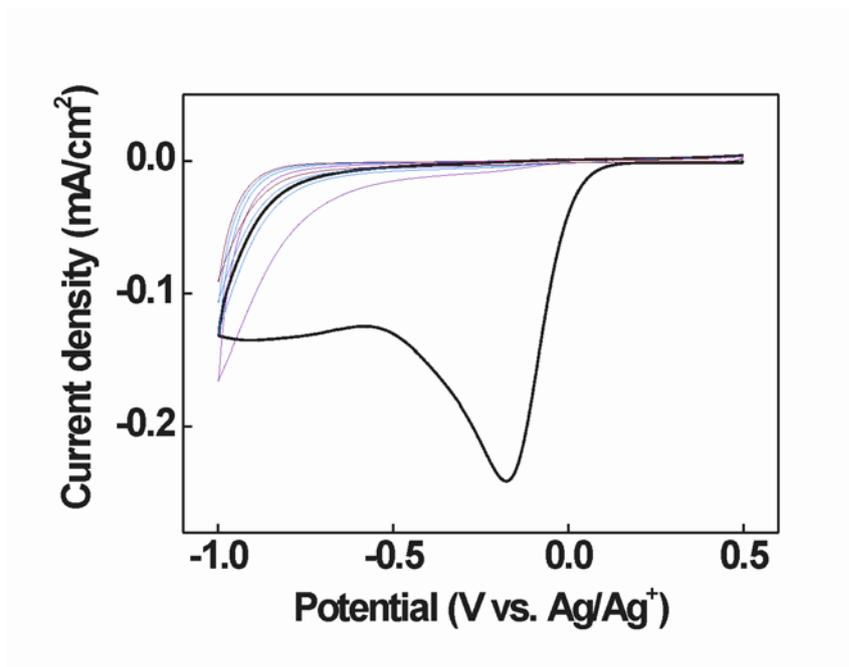

Fig. 29

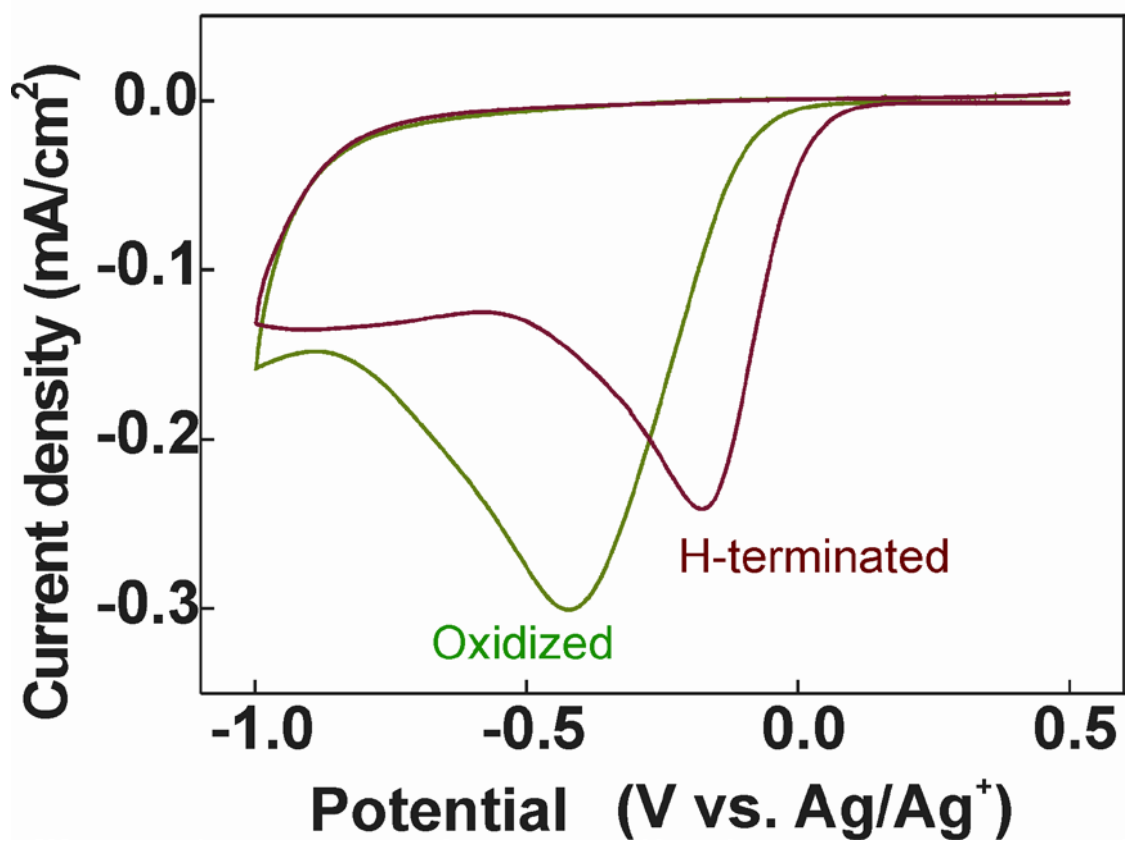

Fig. 30

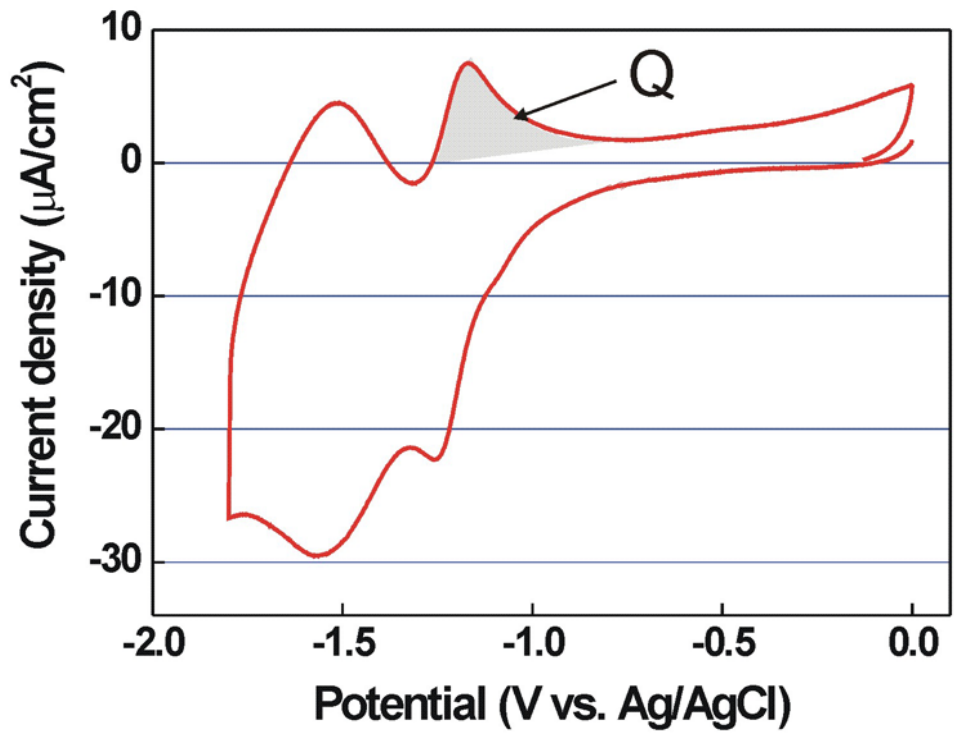

Fig. 31

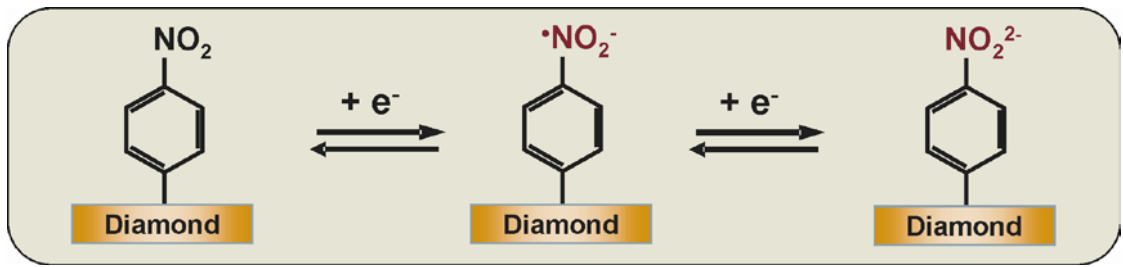

Fig. 32

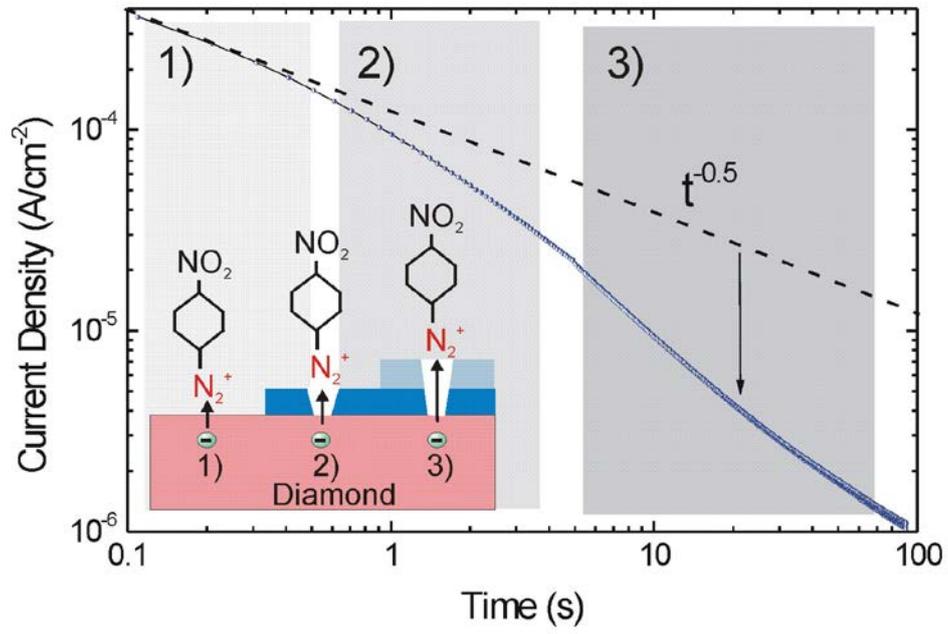

Fig. 33

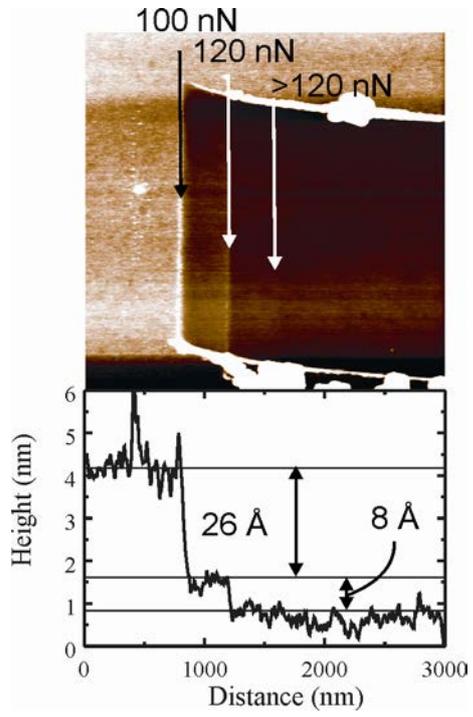

Fig. 34

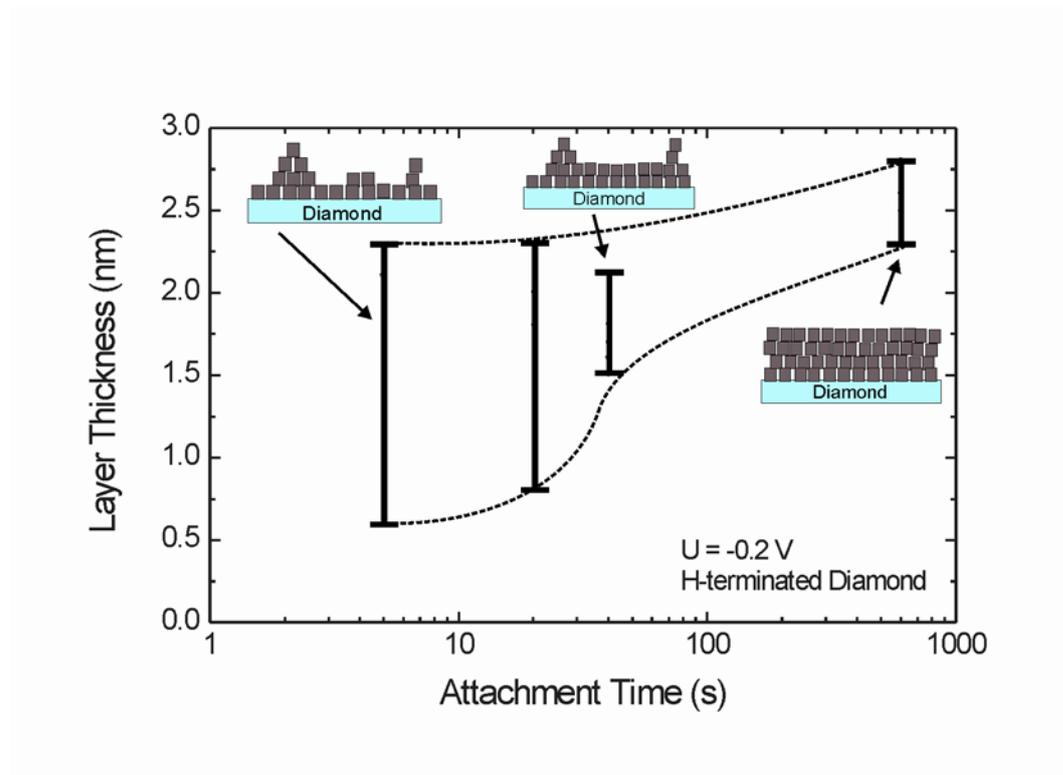

Fig. 35

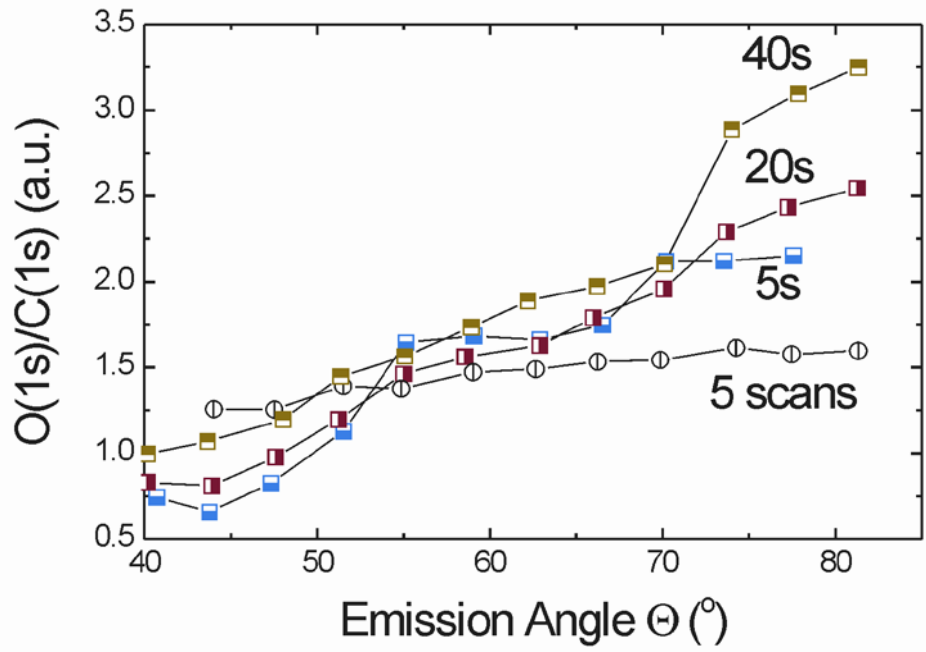

Fig. 36

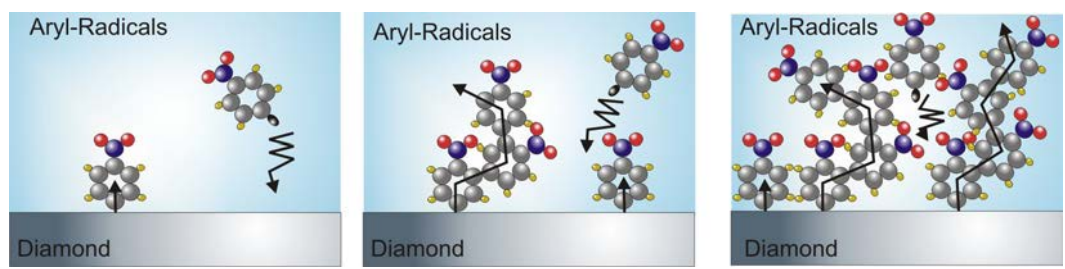

Fig. 37

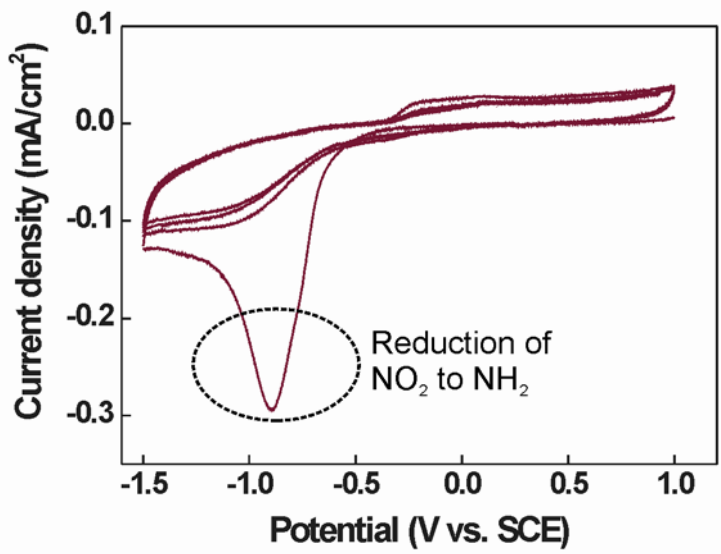

Fig. 38

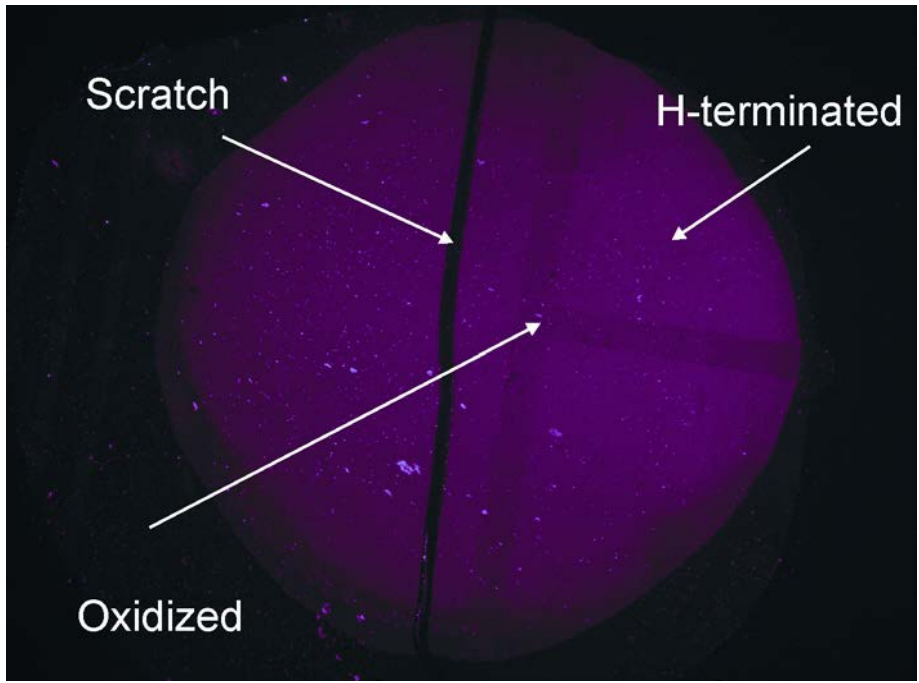

Fig. 39

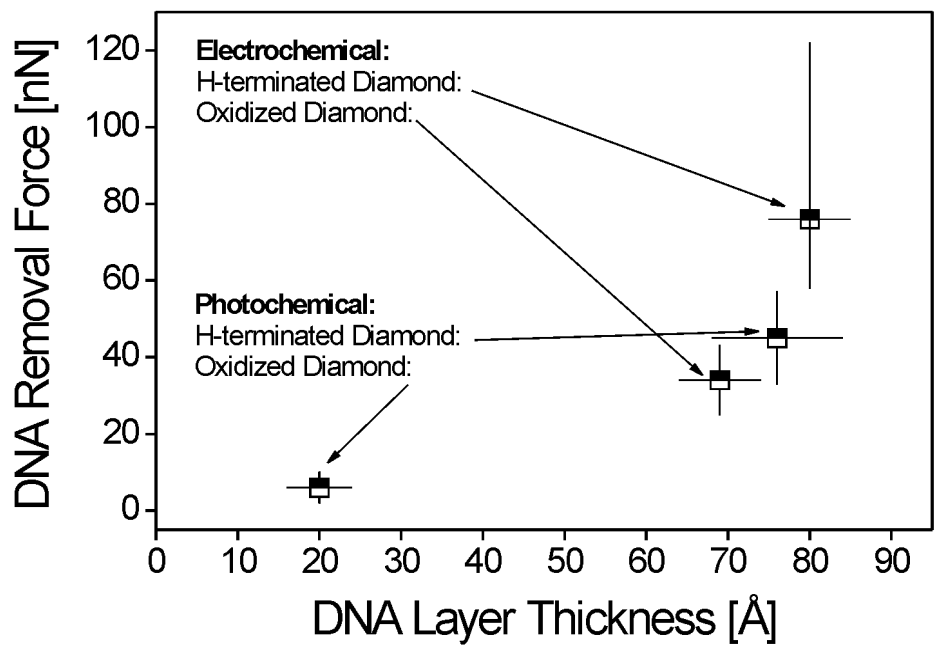

Fig. 40

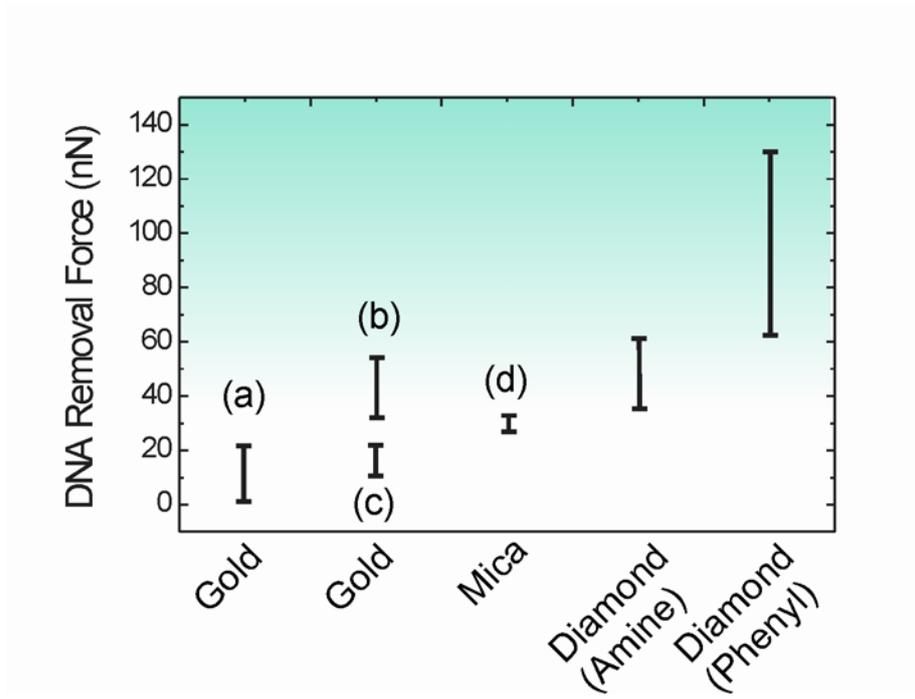

Fig. 41

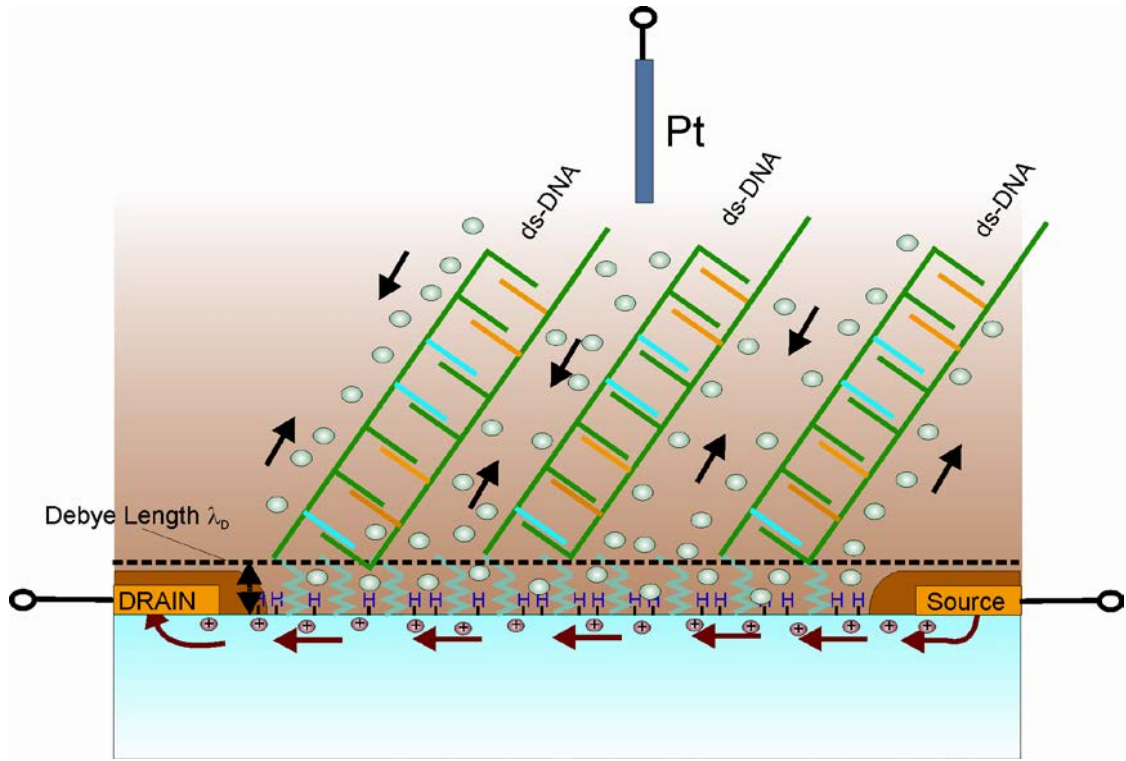

Fig. 42

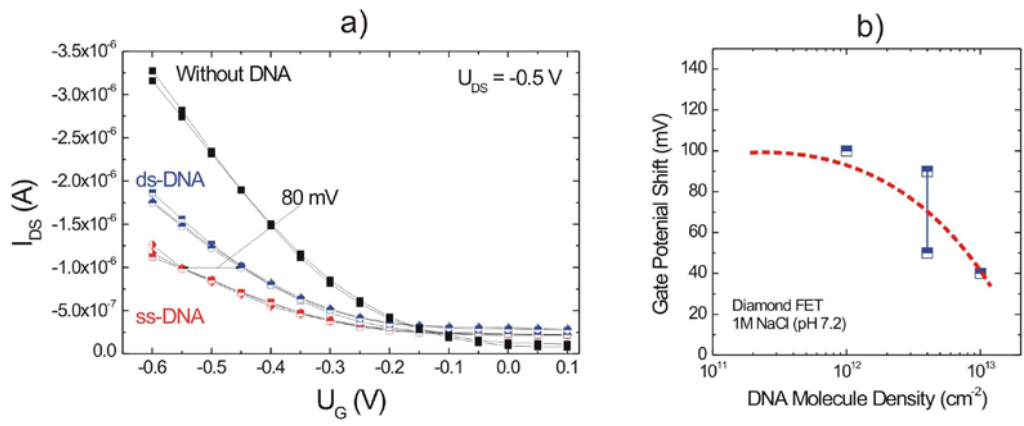

Fig. 43

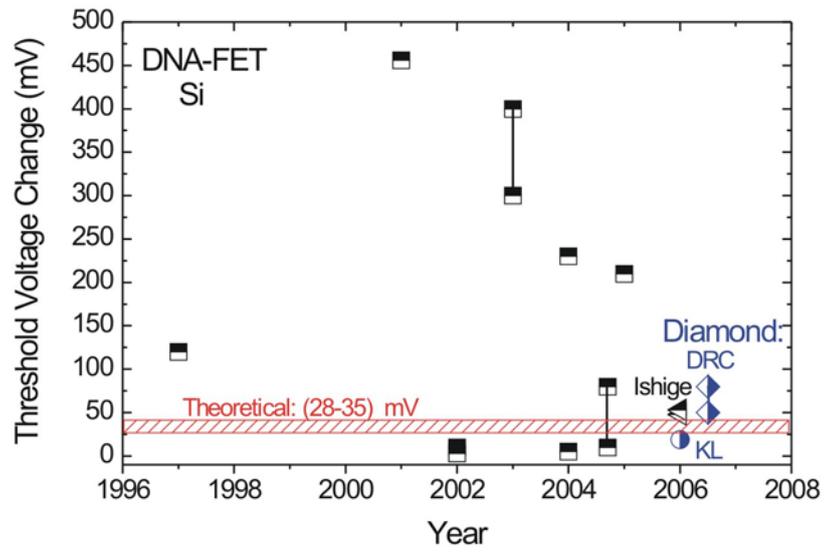

Fig. 44

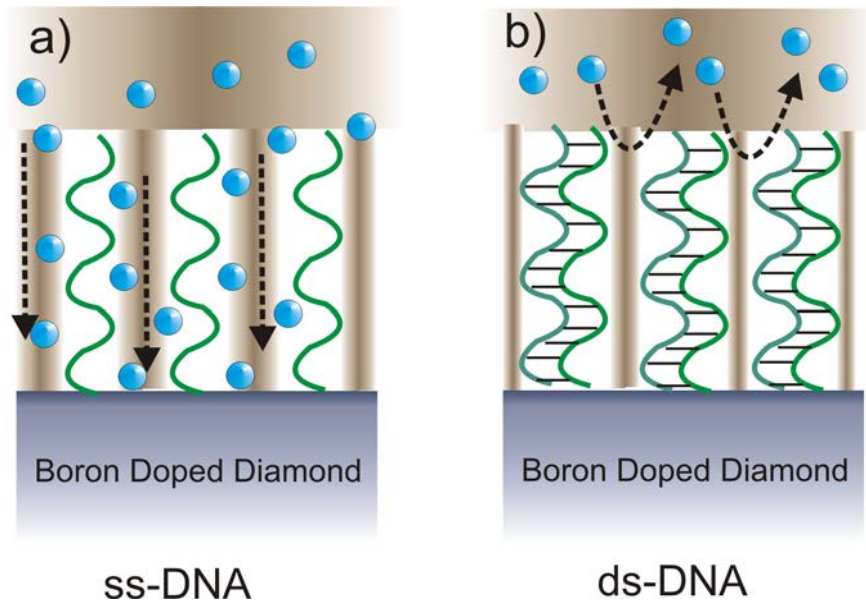

Fig. 45

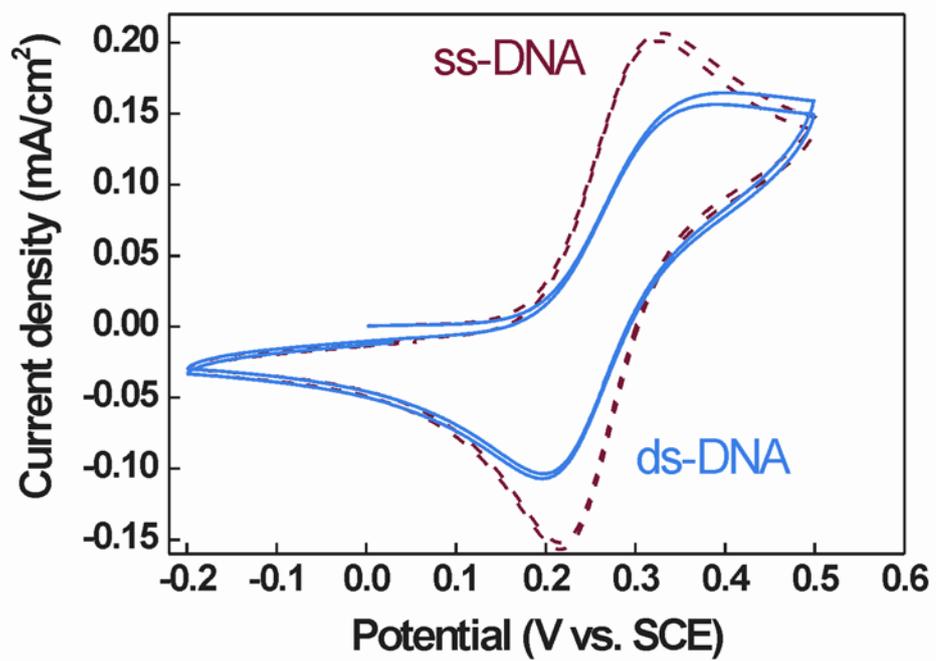

Fig. 46

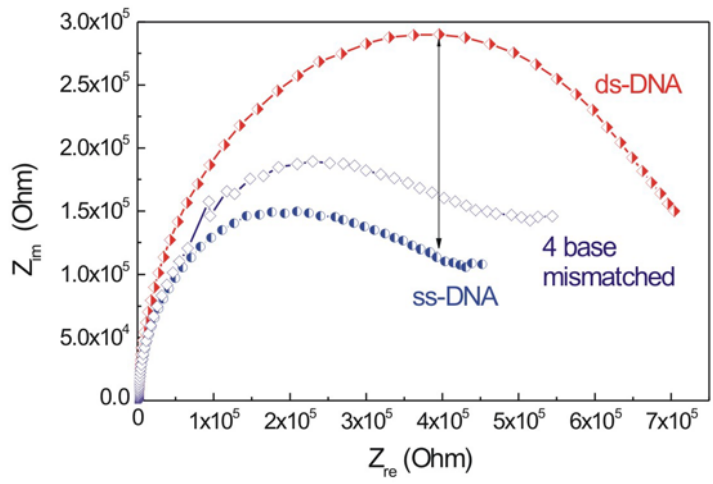

Fig. 47